\documentclass[12pt,twocolumn]{aastex62}

\usepackage{hyperref}
\usepackage{amsmath,mathtools,mathrsfs,amssymb}
\usepackage{graphicx}
\usepackage{bm}
\usepackage[percent]{overpic}
\usepackage{cancel}

\defcitealias{dalpino_lazarian_2005}{GL05}
\defcitealias{dalpino_etal_2010_GPK10}{GPK10}
\defcitealias{kadowaki_etal_15}{KGS15}
\defcitealias{lazarian_vishiniac_99}{LV99}
\defcitealias{xu_lazarian_2023}{XL23}

\defcitealias{Medina-Torrejón_2021}{MGK+21}
\defcitealias{Medina-Torrejón_2023}{MGK23}

\submitjournal{JHEAP}

\usepackage{multirow}

\begin{document}

\title{A Turbulence-Driven Magnetic Reconnection Model for the High-Energy Neutrino Emission from NGC 1068}

\author[0000-0001-6636-4200]{L. Passos-Reis}
\affiliation{Instituto de Astronomia, Geof\'{i}sica e Ci\^{e}ncias Atmosf\'{e}ricas (IAG), Universidade de S\~{a}o Paulo, Rua do Mat\~{a}o 1226, CEP: 05508-090, S\~{a}o Paulo - SP, Brazil.}

\author[0000-0001-8058-4752]{E. M. de Gouveia Dal Pino}
\affiliation{Instituto de Astronomia, Geof\'{i}sica e Ci\^{e}ncias Atmosf\'{e}ricas (IAG), Universidade de S\~{a}o Paulo, Rua do Mat\~{a}o 1226, CEP: 05508-090, S\~{a}o Paulo - SP, Brazil.}

\author[0000-0001-9980-5973]{J. C. Rodríguez-Ramírez}
\affiliation{Centro Brasileiro de Pesquisas Físicas, Rua Dr. Xavier Sigaud 150, CEP: 22290-180, Rio de Janeiro - RJ, Brazil.}

\author[0000-0002-7009-9232]{G. H. Vicentin}
\affiliation{Instituto de Astronomia, Geof\'{i}sica e Ci\^{e}ncias Atmosf\'{e}ricas (IAG), Universidade de S\~{a}o Paulo, Rua do Mat\~{a}o 1226, CEP: 05508-090, S\~{a}o Paulo - SP, Brazil.}

\begin{abstract}
We model the Seyfert II AGN NGC 1068 within a turbulence-induced magnetic reconnection framework to explain its
 high-energy emission. Observations reveal 
a neutrino flux excess higher than the
observed GeV gamma-ray emission by orders of magnitude, with no detected TeV counterpart, suggesting 
efficient hadronic acceleration in the nuclear region with
strong gamma-ray absorption.
Assuming that proton acceleration occurs in a turbulent reconnection layer via a first-order Fermi process, we use a lepto-hadronic model based on a coronal-accretion disk configuration in which magnetic field lines anchored to the $2 \times 10^{7} M_{\odot}$ black hole horizon reconnect with field lines from the inner accretion disk corona. Our model matches the observed spectral energy distribution with
a magnetic field $B_{c} \sim 10^{4}$ G and magnetic reconnection power $\dot{W_{B}} \sim 10^{43}$ erg s$^{-1}$, with $\sim 50\%$ 
efficiency in proton acceleration.
Unlike previous studies, we find that both particle
acceleration and emission take place in the inner region, where protons reach $\sim 10^{14}$ eV via 
first-order Fermi acceleration
within the turbulent reconnection layer, 
rather than drift acceleration.
These protons interact with disk 
photons, coronal X-rays, and 
coronal protons,
producing 
neutrinos, predominantly via $pp$ interactions, at levels consistent with IceCube detections. The associated gamma-rays 
are 
attenuated by $\gamma\gamma$ annihilation, remaining below
current upper limits. Turbulence-driven reconnection is thus a viable mechanism for neutrino production in the coronal region of NGC 1068 and similar sources.
\end{abstract}





\section{Introduction} \label{sec:1_intro}

The nearby active galaxy NGC 1068 is a low-luminous-AGN (LLAGN), and one of the best-studied examples of a Seyfert Type II galaxy. Also known as Messier 77, it has been a key piece for the Active Galactic Nuclei (AGN) Unification Model, where the observed difference between Type I and Type II Seyferts is attributed to the line of sight relative to a central obscuring, dusty torus surrounding the supermassive black hole and its accretion disk. In X-rays, NGC 1068 is classified as a Compton-thick AGN \citep{Ricci2015_comptonthick}, meaning its central engine is heavily obscured by a column density of gas and dust greater than $\sim 10^{24}$ cm$^{-2}$, which absorbs almost all direct nuclear radiation below $\sim 10$ keV, leaving an observed spectrum dominated by reprocessed emission, including a strong neutral Iron K$\alpha$ line \citep{Marinucci2016_nuStar, Zaino2020_nuStar}. Moreover, X-ray monitoring revealed direct nuclear radiation, pointing to a dynamic obscuring environment.

Recent observations have placed NGC 1068 as an important piece to study multi-messenger emission from AGNs, offering a unique window into high-energy particle acceleration in obscured environments. Using 2011–2020 data, the IceCube Collaboration reported evidence for high-energy muon neutrinos from NGC 1068, with an excess of $79^{+22}_{-20}$ events at $4.2\sigma$ significance \citep{IceCube2022}. This signal points to hadronic particle acceleration and, together with the absence of accompanying TeV $\gamma-$rays, suggests strong internal $\gamma\gamma$ absorption in a dense nuclear environment \citep[][see also \citealp{Eichmann2022_ngcmodel}]{Murase2022}.
More recently, a follow-up IceCube study confirmed NGC 1068's status as the most significant neutrino source in the Northern Sky \citep{IceCube2025}. In these proposed disk-corona models, X-ray photons generated in the corona provide the target for proton-photon
interactions that produce neutrinos and may lead to a strong absorption of secondary $\gamma$-rays, offering a consistent picture for the simultaneous presence of high-energy neutrinos and the absence of high-energy gamma-rays.




Motivated by these results, we here explore a lepto-hadronic model in which turbulence-driven magnetic reconnection accelerates particles in the inner region of a dense corona surrounding the supermassive black-hole (BH), incorporating pion production, $\gamma\gamma$ pair production, synchrotron, and inverse-Compton losses.




The proposal that magnetic reconnection can drive particle acceleration and explain the origin of high energy emission in accretion disk-coronal flows around black holes is not new. Several earlier studies have investigated this process ranging from early flare-heated disk theories 
\citep{Liu_Shibata_2002, dalpino_lazarian_2005} 
to more recent investigations of particle acceleration and gamma-ray/neutrino emission \citep[e.g.,][]{dalpino_etal_2010, kadowaki_etal_15, singh_etal_16, kadowaki_etal_2019, khiali_etal_15, Khiali2016,rodriguezramires_etal_18}.


\citet{kadowaki_etal_15}, for instance, plotting the $\gamma $-ray luminosity versus  black hole (BH) mass for several sources, including LLAGNs 
like NGC1068, 
black hole binaries (BHBs), blazars, and GRBs, spanning 10 orders of magnitude in mass and power, obtained 
two distinct correlations:  one for blazars and GRBs, associated with Doppler-boosted jet emission, and another one for LLAGN and BHBs, suggesting a different emission region for their $\gamma$-ray emission. The correlation  obtained  for these sources was found to align well with the magnetic power extracted from turbulent-driven magnetic reconnection  layers formed from the interaction between the BH magnetosphere and accretion disk field lines in the turbulent corona \citep{dalpino_lazarian_2005, dalpino_etal_2010_GPK10, dalpino_etal_2010, kadowaki_etal_15, singh_etal_15, Khiali2016}.  This model is also supported by MHD and GR-MHD simulations of accretion-coronal  flows \citep[e.g.][]{Kadowaki_2018, rodriguezramires_etal_18,kadowaki_etal_2019, dalpino_etal_2020}. 
\citet{Khiali2016}, for instance,  used this model to explain diffuse neutrino emission from  AGNs.

More recent studies, examining the potential origin of NGC 1068’s emission, highlight the difficulty of explaining the observed neutrino peak alongside the absence of VHE gamma rays. Investigations based on energy dissipation driven primarily by plasmoid-mediated (tearing-mode) magnetic reconnection have yielded inconclusive, and in some cases even conflicting results 
regarding the role of reconnection in the acceleration and dissipation processes \citep[e.g.,][]{Mbarek2023, Fiorillo2024a_ApJL, Karavola2025}.


\citet{Mbarek2023} for instance, concluded that protons accelerated in the coronal region by this process cannot account for the observed neutrinos and suggested that they have to be pre-accelerated elsewhere, either in the reconnection layer at the outflow boundary with the disk or in intermittent current sheets in the BH magnetic flux eruptions. \citet{Fiorillo2024a_ApJL}
considered the same acceleration mechanism  and assumed that the protons are linearly accelerated over time via drift in the magnetic field gradients around the current sheets. 
However, since drift is efficient only at low energies, as it depends on particle's energy, they assumed that particles escape time is shorter than the drift acceleration timescale, and then used the escape time to set the size of the acceleration/emission region ($\sim 5 \,R_{\rm g}$, being $R_{\rm g} = GM / c^{2}$) and the maximum proton energy. 
\citet{Karavola2025} considered the same plasmoid-driven scenario as in \citet{Fiorillo2024a_ApJL} and extended the study of neutrinos to other AGNs.
\citet{Fiorillo2024b_ApJ} advanced these ideas by assuming that the large-scale turbulent corona is powered by magnetic reconnection, with the energy release rate set by the reconnection timescale obtained from PIC simulations. A fraction of this energy was then allocated to a small subset of coronal protons (with densities comparable to electrons, as in standard MHD plasmas), which were accelerated stochastically by turbulence outside the current sheets. In this framework, the acceleration time is independent of particle energy, leading to more efficient neutrino production than in earlier models. However, in their treatment, reconnection and turbulence act in separate regions.

It is worth noting that the aforementioned studies \citet{Mbarek2023, Fiorillo2024a_ApJL}
also relied on 3D GRMHD simulations \citep[e.g.,][]{Ripperda2022} as evidence for plasmoid-driven reconnection occurring either in the disk’s equatorial region or at the disk–outflow interface \citep[see also][]{Davelaar2023}. However, recent high-resolution resistive MHD simulations of current sheets \citep{Vicentin2025b, MorilloAlexakis2025} show that $fast$ plasmoid-driven reconnection, with $v_{\rm rec}/v_A \sim 0.01$, should  not occur for the Lundquist numbers (or magnetic resistivities) explored in \citet{Ripperda2022}. Instead, the plasmoids reported in those simulations are likely artifacts of numerical resistivity or simply manifestations of turbulent eddies. In the much higher–resolution MHD simulations of \citet{Vicentin2025b}, the tearing-mode (plasmoid) reconnection rate remains slow and explicitly resistivity-dependent up to very large Lundquist numbers, beyond which the system has already transitioned to a turbulent state.
By contrast, reconnection triggered by MHD turbulence \citep{lazarian_vishiniac_99,eyink2013,Lazarian2020review}, 
does achieve fast, resistivity-independent rates with 
$v_{\text{rec}}/v_A \sim 0.05 - 0.1$, consistent with the theory \citep{kowal_etal_09, takamoto_etal_15, Vicentin2025, Vicentin_2026_PoS}.
This strongly suggests that turbulence-driven reconnection is the dominant mechanism for fast energy release in MHD systems such as AGN accretion flows.

Other possibilities have been
proposed in the literature such as diffusive shock acceleration \citep{Inoue2020}, or gyroresonant stochastic
acceleration \citep{Murase_2020}
to produce neutrino emission in NGC1068. However, the main question that remains open is: what mechanism
would accelerate protons in the corona up to the required
energies 
to produce high-energy neutrinos 
of tens of TeV? 


In this work, we investigate particle acceleration in the coronal accretion flow of the source NGC1068. Instead of plasmoid-driven (tearing mode) reconnection, we adopt a simpler model perspective following \citet{dalpino_lazarian_2005} and \citet{kadowaki_etal_15}, assuming that $turbulence-driven$ magnetic reconnection is the dominant mechanism of hadronic acceleration and subsequent neutrino production,  
in the turbulent corona surrounding the black hole. In contrast to the earlier works of the source mentioned above, turbulence and the reconnection layer are in the same location and the particle acceleration is governed predominantly by a first-order Fermi process inside the reconnection layer \citep{dalpino_lazarian_2005, xu_lazarian_2023}.
In this case, particle energy increases exponentially rather than linearly, enabling efficient particle acceleration with a timescale independent of the particle's energy \citep{kowal_etal_2012, delvalle_etal_16, Medina-Torrejon_2021, Medina-Torrejon_2023, dalpino_medina_2024}.

The contents of the paper are organized as follows. In Section \ref{sec:2_model}, we describe the accretion disk-corona physical scenario, and the key role of turbulence-driven fast magnetic reconnection in particle acceleration, based on the models of \citet{dalpino_lazarian_2005}, 
\citet{dalpino_etal_2010_GPK10}, 
and \citet{kadowaki_etal_15}, as well as the radiative loss timescales for particle interactions and  the spectrum of the accelerated particles. Within this framework, in Section~\ref{sec:3_results} we describe the derived parametric space,  the background coronal radiation field  and the resulting  Spectral Energy Distribution (SED) for NGC 1068. Finally, in Section \ref{sec:4_discussion}, we summarize our findings and discuss their implications for hadronic cascading, high-energy neutrino emission, and the broader context of multimessenger astronomy.

\section{Description of the Model}
\label{sec:2_model}

\subsection{Accretion Disk–Corona Scenario and Fast Magnetic Reconnection Near the Black Hole}
\label{subsec:2.1_scenario}

We adopt the coronal accretion-disk model described in detail by \citet{dalpino_lazarian_2005} \citepalias{dalpino_lazarian_2005} and \citet{kadowaki_etal_15} \citepalias{kadowaki_etal_15} \citep[see also][]{dalpino_etal_2010_GPK10,dalpino_etal_2010,khiali_etal_15,Khiali2016}. This framework assumes a standard geometrically thin and optically thick accretion disk \citep{shakura_sunyaev_73} around the supermassive black hole (SMBH), consistent with the inferred accretion power for the 
LLAGN NGC 1068. Above and below the disk, differential rotation and magnetic field amplification naturally produce a hot, tenuous, and magnetized corona, as supported by analytical and numerical studies \citep[e.g.][and references therein]{Kadowaki_2018}, which serves as the site for magnetic energy dissipation.

\begin{figure}[htbp]
    \centering
    \includegraphics[width=\linewidth]{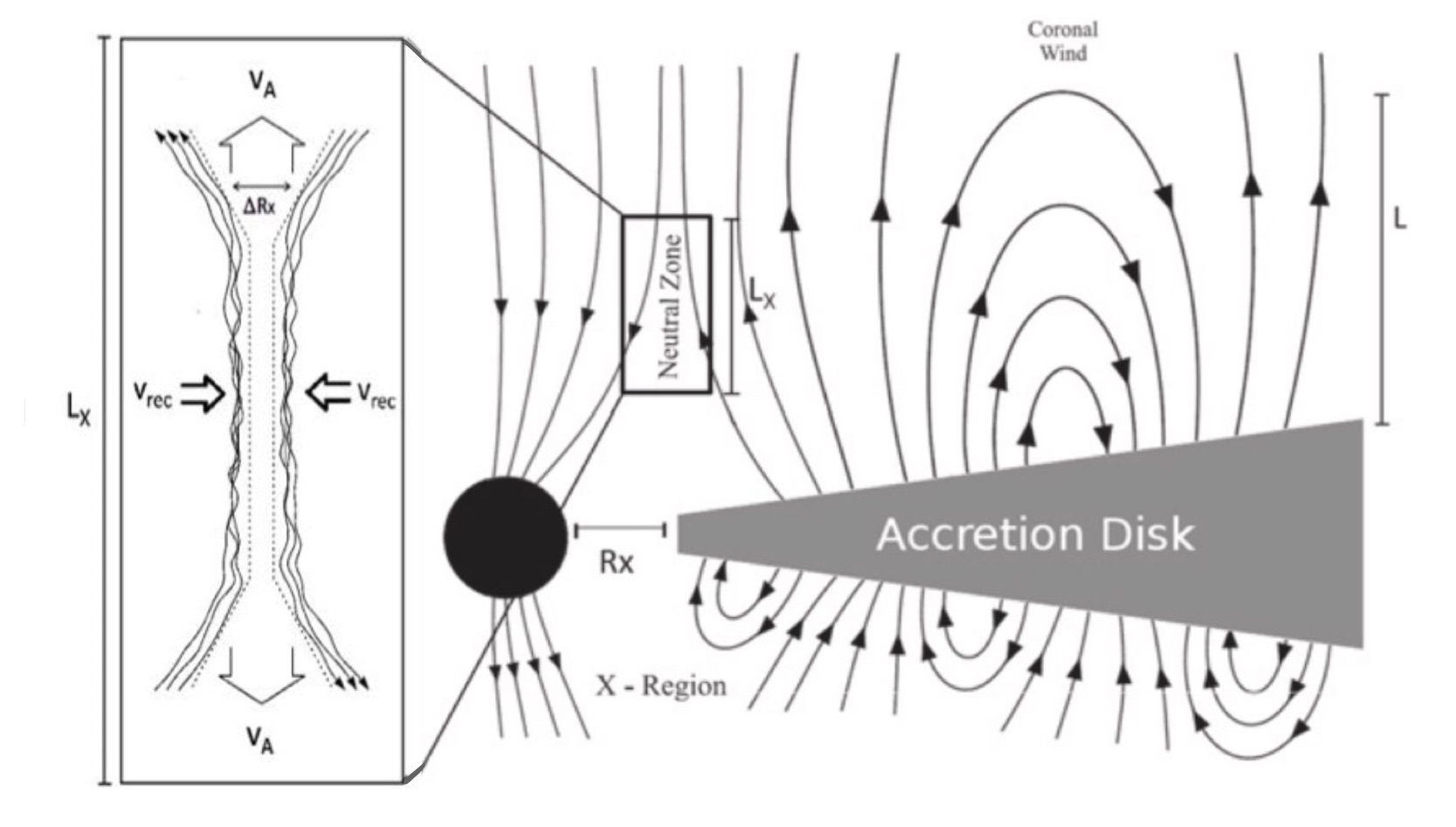}
    \caption{
    Schematics of the magnetic field geometry in the inner accretion flow region, adapted from \citet{dalpino_lazarian_2005, kadowaki_etal_15, kadowaki_etal_2019}. At the inner radius ($R_X$), magnetic field lines arising from the accretion disk into the corona interact with those anchored in the BH magnetosphere. Embedded turbulence drives fast magnetic reconnection \citep{lazarian_vishiniac_99} in the resulting magnetic discontinuity (current sheet) between the two magnetic fluxes of opposite polarity, characterized by a height $L_X$ and a width $\Delta R_X$. The left panel highlights the turbulence-induced wandering of the magnetic field lines, which increases the reconnection rate and boosts particle acceleration  efficiency (see text for details).
    }
    \label{fig:coronal_model}
\end{figure}

Figure~\ref{fig:coronal_model} illustrates the magnetic field configuration where field lines arising from the accretion disk interact and reconnect with those anchored in the black hole (BH) 
horizon.
In this framework, the BH magnetosphere can be maintained by the continuous advection/dragging of magnetic field lines by the accretion flow.
The large-scale poloidal component may arise either from inward advection of field lines from the outer disk or from an in situ disk dynamo, likely driven by
magnetorotational instability (MRI) together with differential rotation \citep[see][and references therein]{dalpino_lazarian_2005,kadowaki_etal_15}. The dynamo action produces periodic magnetic field reversals, giving rise to the emergence of opposite-polarity magnetic fluxes in the inner region, as illustrated in Figure~\ref{fig:coronal_model}.

As in \citet{dalpino_lazarian_2005,dalpino_etal_2010_GPK10, kadowaki_etal_15}, we  
assume that the intensity of the poloidal field that was dragged by the disk and anchored into the BH horizon neighborhood, is of the order of the inner disk magnetic field intensity,
$B_{d}$.  
This is estimated by balancing between the magnetic pressure of the BH magnetosphere and the accretion ram pressure \citep{dalpino_lazarian_2005}, following

\begin{equation}\label{eq:balance_magpress_accrampress}
    \frac{\dot{M}}{4 \pi R^{2}} \left( \frac{2 G M}{R} \right)^{1 / 2} \sim \frac{B_{d}^{2}}{8 \pi }.
\end{equation}

The solution for the intensity of the coronal magnetic field,
$B_c \approx B_d$, at the inner radius $R_X$ (Figure~\ref{fig:coronal_model}), is then given by

\begin{equation}\label{eq:B_d}
    B_{d} \simeq 9.96 \times 10^{8} r_{X}^{-5/4} \dot{m}^{1/2} m^{-1/2} \text{ G},
\end{equation}
where $r_{X} = R_{X}/R_{\rm Sch}$ is the inner disk radius in units of the Schwarzschild radius 
($R_{\rm Sch} = 2GM / c^{2} = 2.96 \times 10^{5} M / M_{\odot } \,\text{cm}$). The black hole mass $m = M / M_{\odot}$ is normalized in solar masses, and $\dot{m} = \dot{M} / \dot{M}_{\rm Edd}$ is the mass accretion rate in units of the Eddington rate ($\dot{M}_{\rm Edd} = 1.45 \times 10^{18} m$ g s$^{-1}$).

The inner radius of the accretion disk, $R_X$, sets the location of the reconnection region. Previous studies \citep[][; \citealt{PassosReis_NGC_ICRC2025}]{dalpino_lazarian_2005, dalpino_etal_2010_GPK10} placed $R_X$ close to the last stable orbit (within $3 \,R_{\rm Sch}$), but doing so can lead to mathematical singularities in the standard Shakura--Sunyaev disk solutions. Moreover, our model does not include full general-relativistic (GR) curvature effects, which may become important for particle and photon trajectories when $R_X<6R_{\rm Sch}$ \citep{kadowaki_etal_15}\footnote{Adopting a pseudo-Newtonian gravitational potential \citep{paczynsky-wiita1980}—which reproduces the leading general relativistic effects near the event horizon while retaining a Newtonian formulation—one can estimate the correction to a purely Newtonian treatment required to account for relativistic effects. This correction is of order unity for $R_X\gtrsim 6R_{\rm Sch}$. For instance, at $R_X=5 R_{\rm Sch}$, the magnetic-field strength computed with the pseudo-Newtonian potential ($B_{d,\rm PSN}$) differs only slightly from the Newtonian value ($B_{d,\rm N}$), with $B_{d,\rm PSN}/B_{d,\rm N}\simeq 1.06$, while the effective free–fall speed of the accretion flow (and thus gravity)  is enhanced by about $12\%$
compared to the Newtonian case. 
}. Therefore, for simplicity and mathematical robustness, we adopt the same threshold as \citet{kadowaki_etal_15}, namely $R_X\simeq 6R_{\rm Sch}$.

In \citet{dalpino_etal_2010_GPK10}, 
Equation \ref{eq:balance_magpress_accrampress} derived an equation set parametrized in terms of the ratio between the total disk pressure (sum of gas and radiation pressures) and the magnetic pressure, denoted by $\beta $. In the case of a radiation-pressure dominated disk \citep{shakura_sunyaev_73}, both parametrizations are related through the equation \citep{kadowaki_etal_15, dalpino_etal_2010_GPK10}:

\begin{equation}\label{eq:beta}
    \beta \simeq 0.12 \ \alpha^{-1} r_{X} \ \dot{m}^{-1},
\end{equation}
where $\alpha$ is the dimensionless Shakura-Sunyaev viscosity parameter.

The surface disk temperature, $T_d$, will define
the blackbody radiation photon field that serves as a target for photohadronic interactions and $\gamma\gamma$ annihilation (see Section \ref{subsec:3.2_radfield_gammaabsorption}). Following the geometrically thin, optically thick, stationary accretion-disk approximation with a Keplerian profile \citep{shakura_sunyaev_73}, the surface 
disk temperature at radius $R_X$ is given by\footnote{We note that   
\citet{dalpino_etal_2010_GPK10} presents the relation for the disk temperature at the midplane. Here, to derive the black-body radiation arising from the optically thick disk
we need the disk surface temperature, however the difference between the two values is approximately a factor of 5.}

\begin{equation}\label{eq:T_d}
   T_{d} \simeq 6.29 \times 10^{7} \, m^{-1/4} \dot{m}^{1/4} r_{X}^{-3/4} q \text{ K}.
\end{equation}

In order to quantify the parameters 
in the hot, magnetized coronal plasma, right above (and below) the inner disk region, we use the relations derived in \citet{kadowaki_etal_15}. The coronal temperature ($T_c$) and number density ($n_c$) at the reconnection site are estimated by \citep{kadowaki_etal_15}:

\begin{equation}\label{eq:T_c}
    T_{c} \simeq 2.73 \times 10^{9} \Gamma^{1/4} r_{X}^{-3/16} l^{1/8} q^{-1} \dot{m}^{1/8} \text{ K};
\end{equation}

\begin{equation}\label{eq:n_c}
    n_{c} \simeq 8.02 \times 10^{18} \Gamma^{1/2} r_{X}^{-3/8} l^{-3/4} q^{-2} \dot{m}^{1/4} m^{-1} \text{ cm}^{-3}.
\end{equation}

The factors, $\Gamma $ and $q$, account for special-relativistic effects on the Alfvén speed and geometry correction of the inner disk, respectively, and are given by

\begin{equation}
    q = \left[ 1 - \left(\frac{3 R_{\rm Sch}}{R_{X}} \right)^{\frac{1}{2}} \right]^{\frac{1}{4}} \text{and} \ \Gamma = \left(1 + \left( \frac{v_{A0}}{c} \right)^{2} \right)^{-\frac{1}{2}},
\end{equation}


\noindent
with $v_{A0} = \frac{B}{\sqrt{4 \pi \rho}}$. The model uses the proton rest mass ($m_H$) and a mean molecular weight of $\mu \sim 0.6$, so that $\rho_c = \mu m_{H} n_{c}$.

The 
width of the recconection region, $\Delta R_{X}$, placed in $R_X$, can be much smaller than the height of the coronal region $L$, and the extension height of the reconnection region (turbulent neutral zone), $L_{X}$, considered in Figure \ref{fig:coronal_model}, but the presence of turbulence enlarges this region \citep{lazarian_vishiniac_99}.
In \citet{lazarian_vishiniac_99}, 
the field lines are advected out of  the reconnection layer of length $L_X$, at a velocity of a fraction of $v_A$, which depends only on the turbulence parameters
and not on the magnetic
resistivity.
The wandering of magnetic field lines caused by the turbulence determines 
this reconnection rate as \citep{lazarian_vishiniac_99}:

\begin{equation}
    v_{\rm rec} = v_A \min \left[ \left( \frac{L_X}{\ell} \right)^{1/2}, \left( \frac{\ell}{L_X} \right)^{1/2} \right] \left( \frac{v_{\ell}}{v_A} \right)^2
\end{equation}
where $\ell$ and $v_{\ell}$ are the injection scale of the turbulence and its velocity at this scale, respectively.
This limit on the reconnection speed is fast, both in the sense that it does not depend on the resistivity, and in the sense that it represents a large fraction of the Alfvén speed.

And the corresponding width, $\Delta R_X$, is directly determined from its relation with $v_{\rm rec}$:

\begin{equation}
    v_{\rm rec} = \left( \frac{\Delta R_X}{L_X} \right) v_A 
\label{vrec_def}
\end{equation}
where $v_A = \Gamma v_{A0}$.

Using the equations above for turbulence-induced reconnection and  the parametrization of our model, $\Delta R_X$ is given by \citep{kadowaki_etal_15, kadowaki_etal_2018b}: 

\begin{equation}\label{eq:width_rec}
    \Delta R_{X} \simeq 11.6 \Gamma^{-5/4} r_{X}^{31/16} l^{-5/8} l_{X} q^{-3} \dot{m}^{-5/8} m \text{ cm}.
\end{equation}

The efficiency of particle acceleration depends on the magnetic power released through reconnection. Following \citet{dalpino_lazarian_2005, kadowaki_etal_15}, we find that the magnetic reconnection power produced by turbulence-driven fast reconnection in the vicinity of the black hole, at $R_X$, can be expressed as:


\begin{equation}\label{eq:reconnec_power}
    \dot{W_{B}} = 1.66 \times 10^{35}\ \Gamma^{-1/2}\ r_{X}^{-5/8}\ l^{-1/4}\ l_{X}\ q^{-2}\ \dot{m}^{3/4}\ m.
\end{equation}
We assume that a fraction ($\eta_{\rm p}$) of this power is channeled into accelerating protons and the remaining may power the coronal X-rays, though the latter is a  minor contribution to the entire corona, 
as detailed in Section~\ref{subsec:2.4_powerspectrum_partic}.

\subsection{Acceleration Timescale}
\label{subsec:2.2_acc_timescale}

It is generally accepted that particles are accelerated by fast magnetic reconnection primarily via first-order Fermi process occurring inside the reconnection layers, where trapped particles undergo head-on interactions with converging  magnetic fluctuations on both sides of the current sheet \citep{dalpino_lazarian_2005}. This process implies a temporal exponential growth of particle energy and thus, an acceleration time independent on the particle's energy. 
This process and the independence of the acceleration time on the particle's energy have been successfully tested in 3D MHD simulations with turbulent-driven fast reconnection \citep[e.g.,][]{kowal_etal_2012, delvalle_etal_16, Medina-Torrejon_2021, Medina-Torrejon_2023, dalpino_medina_2024}.

In recent work, \cite{xu_lazarian_2023} revisited 
\citet{dalpino_lazarian_2005}'s model, and derived  three different conditions for the particle acceleration time within the turbulent-induced current sheets, depending on the reconnection velocity, the thickness of the reconnection layer, and the angle between the magnetic guide field and the reconnection field. The fastest acceleration time is given by:

\begin{equation}\label{eq:fermi_acc}
    t_{\rm acc, Fermi} \approx \frac{4 \Delta R_{X}}{c d_{\rm ur}},
\end{equation}

where

\begin{equation}
    d_{\rm ur} \approx \frac{2 \beta_{\rm in} \left( 3 \beta_{\rm in}^{2} + 3 \beta_{\rm in} + 1 \right)}{3 \left( \beta_{\rm in} + \frac{1}{2} \right) \left( 1 - \beta_{\rm in}^{2} \right) }.
\end{equation}

This relation has been also successfully tested numerically in recent work \citep{dalpino_medina_2024}. Being $\Delta R_X$ the reconnection layer thickness  (Eq.~\ref{eq:width_rec}), and $\beta_{\rm in} = v_{\rm rec} / c$ the inflow (or reconnection) velocity (Eq.~\ref{vrec_def}) normalized by the light speed.

When a particle’s Larmor radius becomes comparable to the width of the reconnection sheet ($\Delta R_{X}$), it escapes the layer and can undergo additional acceleration in the gradients of the unreconnected magnetic field through drift acceleration \citep[e.g.,][]{kowal_etal_2012}. Because this mechanism scales linearly with the particle’s energy, it becomes inefficient at high energies \citep[e.g.,][]{delvalle_etal_16, dalpino_medina_2024}. As we will show in Section~\ref{subsec:2.3_rad_loss_time}, in the case of NGC 1068, protons are accelerated by the Fermi process within the turbulent reconnection layer and radiate away their energy in the immediate surroundings up to a threshold well below the energy at which drift acceleration becomes relevant.

\subsection{Radiative Loss Timescales}
\label{subsec:2.3_rad_loss_time}

The maximum energy a particle can attain is determined not only by the constraints of the acceleration mechanism (the Larmor radius limit, $r_L \le \Delta R_X$), but also by the competition between the particle acceleration timescale ($t_{\rm acc}$) and the various cooling/loss timescales ($t_{\rm loss}$). In highly magnetized, dense environments, like the corona of an Active Galactic Nucleus (AGN), particle energy losses often dominate, leading to a maximum energy that is cooling-limited rather than Larmor-radius-limited.

We need to compare the constant acceleration rate, provided by the first-order Fermi process within the reconnection layer ($t_{\rm acc, fermi}$, Eq.~\ref{eq:fermi_acc}), against the sum of the relevant energy loss and escape channels.

In this work, we  consider energy loss through four major channels which are mostly due to accelerated protons: Synchrotron cooling ($t_{\rm sync}$), proton-proton collisions ($t_{\rm pp}$), and two photo-hadronic processes: photopion production ($t_{\rm p\gamma}$) and Bethe-Heitler pair production ($t_{\rm BH}$). The total energy loss rate is the sum of the inverse timescales for each process, 

\begin{equation}
    \frac{1}{t_{\rm loss}(E_p)} = \frac{1}{t_{p\gamma }(E_p)} + \frac{1}{t_{pp}(E_p)} + \frac{1}{t_{syn}(E_p)} + \frac{1}{t_{\rm BH} (E_p)},
    \label{loss-rate}
\end{equation}
where $E_p$ is the proton energy. These rates depend on the ambient magnetic field strength, the coronal plasma density, and the structure of the multi-component background photon field (disk blackbody and X-ray corona). The specific mathematical formulations and parameters used for calculating these characteristic timescales are provided in the Appendix~\ref{sec:ap_timescales} \citep[see also][]{rodriguez_ramirez_etal2020, dalpino_rodriguez_txs_2025}.

\subsection{Power Spectrum of the Accelerated Particles}
\label{subsec:2.4_powerspectrum_partic}

As described, a fraction ($\eta_{\rm p}$) of the magnetic reconnection power ($\dot{W_{B}}$) 
goes into accelerating protons, and the remaining can power the coronal X-rays, though the latter is a minor contribution to the entire corona.
The fraction, $\eta_{\rm p}$, is assumed as a free parameter of the model.  

To model the electromagnetic cascade, we assume that the accelerated protons by the turbulence-driven reconnection (Fermi) process, develop a power law injection spectrum $Q(E_p) \propto E_p^{- \alpha_p }$, with spectral index $\alpha_p$.

The corresponding steady-state distribution results
from the balance between
particle injection and total energy losses, which here we approximate as

\begin{equation}
    N_p (E_p) \approx Q(E_p) t_{\rm loss} (E_p),
\end{equation}
where $t_{\rm loss} (E_p)$ is given by 
Eq.~\ref{loss-rate}.

The protons 
undergo 
hadronic interactions, primarily proton-proton (p-p), which produce high-energy neutrinos and gamma-rays through the decay of pions ($\pi^{0}$ and $\pi^{\pm }$). 
The rate of 
$\gamma $-ray production is given by:

\begin{equation}
    \dot{N}_{\gamma } (E_{\gamma }) \propto \frac{N(E_{p})}{\tau_{\rm int} (E_{p})},
\end{equation}
where $\tau_{\rm int} (E_{p})$ denotes the
characteristic timescale of the $\gamma $-ray producing channels under consideration. In particular, $\tau_{\rm int} = t_{pp}$ for proton–proton interactions and $\tau_{\rm int} = t_{p\gamma}$ for photopion production (see Appendix~\ref{sec:ap_timescales}). 


Secondary  electrons and positrons will be produced via Bethe-Heitler, the decay of charged pions, and $\gamma-\gamma$ absorption (Breit-Wheeler) processes) (see also Sections \ref{subsec:2.3_rad_loss_time} and \ref{subsec:3.2_radfield_gammaabsorption}, and Appendix \ref{sec:ap_timescales}), initiating the development of the electromagnetic cascades. 

To calculate the 
emission produced from the secondary leptons, we model the energy distribution $N_{e} (E_{e})$ of these leptons (in units of erg$^{-1}$ cm$^{-3}$) as a stationary solution of the transport equation \citep[e.g.,][]{Ginzburg_Syrovatskii_1964}, as described in \cite{dalpino_rodriguez_txs_2025}:

\begin{equation}\label{eq:N_e}
    N_{e} (E_{e}) = \left| P_{e} \right|^{-1} \int_{E_{e}}^{\infty} d E_{e}' \Phi_{e} (E_{e}').
\end{equation}

The factor $P_{e}$ in the equation above is the total rate of $e^{\pm }$ energy losses due to Synchrotron radiation and Inverse Compton (IC) scattering:

\begin{equation}
    P_{e} = P_{\rm syn} + P_{\rm IC}.
\end{equation}

The corresponding 
flux given by the function $\Phi_{e}$ within the integral of Eq.~\ref{eq:N_e}, is the
electron production rate in units of erg$^{-1}$ s$^{-1}$ cm$^{-3}$ from muon decay, Bethe-Heitler, or Breit-Wheeler processes, 
that we compute following the expressions given in \citet{rodriguez_ramirez_etal2020}, \cite{dalpino_rodriguez_txs_2025}, and references therein.

\section{Results for NGC 1068}
\label{sec:3_results}

\subsection{Model and Observational Parameters}
\label{subsec:3.1_params}


The black hole, accretion disk, and coronal parameters required for our model are presented in Tables~\ref{table_observations} and \ref{table_parameters}.

Table~\ref{table_observations} provides a summary of the parameters inferred from observations, such as the black hole mass $M=2 \times 10^{7} \,M_{\odot}$ \citep{Murase2022} and an accretion rate $\dot{M} = 0.55 \,\dot{M}_{\rm Edd}$ for the Seyfert II galaxy NGC 1068, consistent with values derived from observational constraints 
\citep[e.g.][]{Marinucci2016_nuStar}. It also includes the estimated distance of $d \simeq 10.1 \,\text{Mpc}$ and key measured luminosities:
the $\gamma $-ray luminosity $L_{\mathrm{\gamma-ray}} \simeq (1.70 \pm 0.32) \times 10^{41} \,\text{erg} \,\text{s}^{-1}$, and the bolometric luminosity often inferred to be $L_{\mathrm{bol}} \sim 10^{45} \,\text{erg} \,\text{s}^{-1}$. We note that 
these luminosities are included only as reference measures of the energetics of the system,
rather than being used to constrain model parameters such as the magnetic reconnection power.

\begin{table*}[htbp]
    \centering
    \caption{Key physical properties of NGC 1068 as
    derived from observations.}
    \begin{tabular}{lcc}
    \hline
    \hline
    Parameter & Value & Reference \\
    \hline
    $M$ (Black hole mass) & $2 \times 10^7 \, M_\odot$ & \citet{Woo_Urry_2002, Panessa_2006, Murase2022} \\
    $\dot{M}$ (Accretion rate) & $0.55 \, \dot{M}_{\rm Edd}$ & \cite{Marinucci2016_nuStar} \\
    $d$ (Distance) & $10.1 \pm 1.8 \, \rm{Mpc}$ & \cite{Tully_etal_2009, Salvatore2024_jet} \\
    $z$ (Redshift) & $0.003793$ & \cite{Khachikian_Weedman_1974} \\
    \hline
    $L_{\mathrm{X-ray}}$ (X-ray luminosity) & $10^{43}-10^{44} \,\text{erg} \,\text{s}^{-1}$ & \cite{Marinucci2016_nuStar} \\
    $L_{\mathrm{\gamma-ray}}$ ($\gamma$-ray luminosity) & $(1.70 \pm 0.32) \times 10^{41} \,\text{erg} \,\text{s}^{-1}$ & \cite{Lenain2010} \\
    $L_{\mathrm{bol}}$ (Bolometric luminosity) & $ \sim 10^{45} \,\text{erg} \,\text{s}^{-1}$ & \cite{Woo_Urry_2002, Marinucci2016_nuStar, Murase2022} \\
    \hline
    \end{tabular}
    \label{table_observations}
\end{table*}

\begin{table*}[htbp]
    \centering
    \caption{Adopted parameters within our framework and the respective model-derived parameters at radius $R_X$, based on Eqs.~\ref{eq:balance_magpress_accrampress}–\ref{eq:reconnec_power}, utilized for computing the cooling rates and the SED of NGC\,1068 in the following sections.}
    \begin{tabular}{lccc}
    \hline
    \hline
    This Work & Parameter & Value & Unit \\
    \hline
    Coronal magnetic flux tube height & $l$ & 30 & [$L/R_{\rm Sch}$] \\
    Height of reconnection region & $l_X$ & 30 & [$L_X/R_{\rm Sch}$] \\
    Inner radius of disk & $r_X$ & 6 & [$R_X/R_{\rm Sch}$] \\
    Accretion disk viscosity & $\alpha $ & 0.1 & -- \\
    Spectral index of the injected proton distribution & $\alpha_{\rm p}$ & 1.7 & -- \\
    Accretion power conversion efficiency to coronal X-ray luminosity & $\eta_{\rm cx} $ & 0.0085 & -- \\
    Fraction of reconnection power transferred to protons   & $\eta_{\rm p} $ & 0.5 & -- \\
    \\
    \hline
    \hline
    Model-derived Parameters & & & \\
    \hline
    Coronal magnetic field & $B_c$ & $1.8 \times 10^{4}$ & [G] \\
    Coronal particle density & $n_c$ & $2.1 \times 10^{10}$ & [cm$^{-3}$] \\
    Coronal temperature & $T_c$ & $3.4 \times 10^{9}$ & [K] \\
    Surface disk temperature & $T_d$ & $1.5 \times 10^{5}$ & [K] \\
    Width of current sheet & $\Delta R_{X}$ & $1.5 \times 10^{11}$ & [cm] \\
    \quad \quad in Schwarzschild radii & & $=0.025 \,R_{\rm Sch}$ & [$R_{\rm Sch}$] \\
    Reconnection power released & $\dot{W}_{B}$ & $1.9 \times 10^{43}$ & [erg\,s$^{-1}$] \\
    Reconnection velocity & $v_{\rm rec}$ & 0.001 & [$v_{\rm A}$] \\
    Proton power ($\eta_{\rm p} \dot{W}_{B}$) & $L_{p}$ & $9.7 \times 10^{42}$ & [erg s$^{-1}$] \\
    \hline
    \hline
    \end{tabular}
    \label{table_parameters}
\end{table*}

In our modeling, we set the dimensionless Shakura-Sunyaev viscosity parameter to $\alpha = 0.1$, a typical value for geometrically thin disk models, which, for the $\dot{m}$ parameter adopted in Table \ref{table_parameters} and assuming $R_{X} \simeq 6 \,R_{\rm Sch}$, as discussed in Section~\ref{sec:1_intro} and in \citet{kadowaki_etal_15}, yields the the gas-plus-radiation  to magnetic pressure ratio $\beta \simeq 13$ for the disk, through Eq.~\ref{eq:beta}. Other assumptions in our model parameters include the proton injected  power-law spectrum index $\alpha_{\rm p}$, the coronal 
height ($L$), reconnection region length ($L_X$), as well as the efficiency for the coronal X-ray luminosity ($\eta_{\rm cx}$) (defined in Section~\ref{subsec:3.2_radfield_gammaabsorption}), and the reconnection power fraction used to accelerate the proton   ($\eta_{\rm p}$), totalizing seven free parameters in our model.

Table~\ref{table_parameters} provides both the primary input parameters, 
and the resulting disk-coronal parameters estimated from the set of Equations \ref{eq:balance_magpress_accrampress}–\ref{eq:reconnec_power} at the position $R_X$. Our best fit model for NGC1068 yields a coronal magnetic field $B_{d} \simeq B_{c} \sim 1.8 \times 10^{4}$ G (Eq.~\ref{eq:B_d}) and a total reconnection power of $\dot{W}_{B} \sim 1.9 \times 10^{43}$ erg s$^{-1}$ (Eq.~\ref{eq:reconnec_power}), where the reconnection occurs across a width of $\Delta R_X = 0.025 \,R_{\rm Sch}$ (Eq.~\ref{eq:width_rec}). More parameters include the surface disk temperature $T_d$ (Eq.~\ref{eq:T_d}), the coronal temperature $T_c$ (Eq.~\ref{eq:T_c}), the particle number coronal density $n_c$ (Eq.~\ref{eq:n_c}) and the reconnection velocity (Eq.~\ref{vrec_def}).

\subsection{Background Radiation Fields \& $\gamma $-Ray Absorption Through Photon-Photon Annihilation}
\label{subsec:3.2_radfield_gammaabsorption}

The acceleration region is embedded in a dense radiation field, above and below the accretion disk, which provides the target photons for photohadronic processes ($p\gamma$) and is responsible for the TeV $\gamma$-rays suppression in high-energies, via photon-photon annihilation ($\gamma\gamma$). 

The volume  of the emission region in our model is taken as the spherical region (of volume $V= 4\pi L^3/3$) that encompasses the cylindrical shell around the BH where turbulence-driven magnetic reconnection particle acceleration takes place in Figure \ref{fig:coronal_model}, above and below the accretion disk.

We consider two main background photon fields for the core of NGC 1068:
(i) a blackbody component from the optically thick accretion disk with a surface temperature, $T_d(R)$, based on a 
\citet{shakura_sunyaev_73} disk, 
integrated radially from the inner disk radius ($R_X$) 
up to the outer disk radius $R = L$ (see Eq.~\ref{eq:T_d} and Table~\ref{table_parameters} for the temperature at $R = R_X$), which serves as a proxy for the optical, ultraviolet (OUV) background, and (ii) an X-ray non-thermal coronal photon field constrained by NuSTAR and XMM–Newton observations of NGC 1068 \citep{Marinucci2016_nuStar}.

\begin{figure}[htbp]
    \centering
    \includegraphics[width=\linewidth]{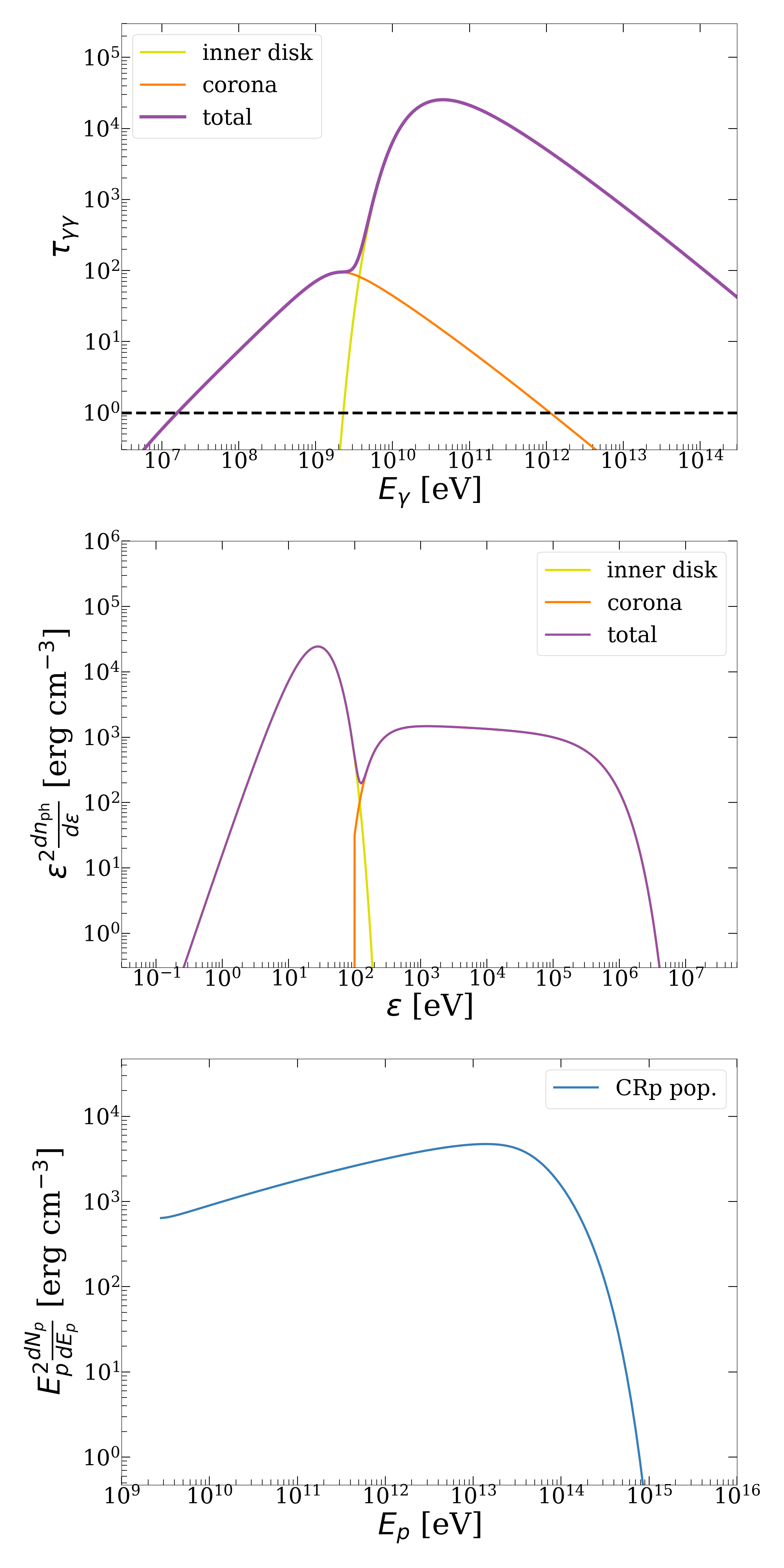}
    \caption{Gamma-ray annihilation opacity, photon energy densities, and and cosmic-ray proton distribution
    in the disk-corona environment.
    \textit{Top panel:} Gamma-ray optical depth ($\tau_{\gamma\gamma}$) as a function of gamma-ray energy ($E_\gamma$) for the accretion-disk blackbody radiation field (yellow curve) and the coronal X-ray photon field (orange curve), along with their combined total opacity (purple curve). \textit{Middle panel:} Photon energy densities of the disk blackbody (yellow) and coronal X-ray (orange) radiation fields,
    shown as $\epsilon^{2} dn_{\rm ph}/ d\epsilon $, in units of erg cm$^{-2}$, as a function of
    photon energy (eV). \textit{Bottom panel:} Accelerated cosmic-ray proton spectrum (\textit{CRp pop.}; blue curve), assuming
    a power-law index $\alpha_p=1.7$. The suppression of $\gamma$-rays for $E_\gamma \gtrsim 100$ GeV is evident in the top panel.}
    \label{fig:tau_gg}
\end{figure}

The coronal non-thermal X-ray emission is modeled by assuming a fraction $\eta_{\rm cx}$ of the accretion power (Table~\ref{table_parameters}), constrained by observations, adopting a single broken power-law spectrum with a peak energy ($E_0 = 200$ eV), a cutoff energy ($E_{\rm cut} = 5 \times 10^{5}$ eV), and a spectral index ($\alpha_X = 2.05$) \citep{Marinucci2016_nuStar}. The corresponding  accretion power is estimated from $\dot{M}_{\rm acc} = \dot{m} \dot{M}_{\rm Edd}$, with $\dot{M}_{\rm Edd} = L_{\rm Edd} M / (\eta_r c^{2})$, where  $\eta_{\rm r} = 0.1$ is the radiative efficiency \citep{Frank_King_Raine_2002}. Thus, the X-ray emission power estimated in our model is $L_{\rm cx} = \eta_{\rm cx} \dot{M}_{\rm acc} c^{2}$,
determined by the fraction $\eta_{\rm cx}$ of the accretion power listed in Table~\ref{table_parameters}. This is consistent with the observed X-ray coronal luminosity, $L_{\mathrm{X-ray}} = 10^{43}$ -- $10^{44}$ erg s$^{-1}$, inferred by \citet{Marinucci2016_nuStar} in the $2$ -- $10$ keV energy band.
 
We note that most of this X-ray coronal emission could be, in principle, powered by magnetic energy released through reconnection processes occurring throughout the entire corona. In particular, the magnetic reconnection power $\dot{W_{B}}$  evaluated in Eq.~\ref{eq:reconnec_power}, which refers exclusively to the reconnection layer located in the inner coronal region at $R_X$, as illustrated in Figure~\ref{fig:coronal_model}, contributes only a fraction of the total coronal power budget.
Since a fraction $\eta_p$ of $\dot{W_{B}}$ is channeled into particle acceleration within the reconnection site (Section~\ref{subsec:3.1_params}, Table~\ref{table_parameters}), the remaining contribution to $L_{cx}$ is  $(1-\eta_p) \dot{W_{B}}$.


Gamma-rays produced in hadronic interactions can be absorbed by the ambient radiation field through electron-positron pair production via the Breit-Wheeler process \citep{Breit_Wheeler_1934}. The interaction of the high-energy $\gamma$-rays ($\gamma$) produced by neutral pion ($\pi^0$) decay with the low-energy target photons ($\gamma_{\rm low}$) leads to electron-positron pair production ($\gamma + \gamma_{\rm low} \rightarrow e^{+} + e^{-}$). This process, known as photon-photon annihilation, results in a high optical depth, $\tau_{\gamma\gamma}$, which attenuates the very-high-energy (VHE) $\gamma$-ray flux. The probability of this absorption is quantified by the optical depth, $\tau_{\gamma\gamma}(E_\gamma)$, which depends on the ambient photon field and the pair production cross-section. The optical depth is given by \citep{Romero_Vila_2008, Romero2010}:

\begin{equation}\label{eq:tau_gg}
    \tau_{\gamma\gamma}(E_\gamma) = \int_{R_{\rm ph}} \int_{\epsilon_{\rm min}}^{\infty} n(\epsilon ) \sigma_{\gamma\gamma}(\epsilon, E_\gamma) \ d\epsilon \ dR,
\end{equation}
where $E_\gamma$ is the gamma-ray photon energy, $\epsilon$ is the target photon energies, 
 $n(\epsilon)$ is the photon number density at energy $\epsilon$, 
$\epsilon_{\rm min} = m_{e}^{2} c^{4} / E_{\gamma }$, and $R_{\rm ph}$ defines the 
length of the path that $\gamma $-rays photons travel before leaving the system.
The absorption condition $\tau_{\gamma\gamma}(E_\gamma) \gg 1$ for TeV energies is essential to explain the lack of $\gamma$-ray counterpart, as shown in the \textit{top panel} of Figure~\ref{fig:tau_gg}.




The pair production cross-section, $\sigma_{\gamma\gamma}$, governs the likelihood of the interaction and is given by \citep{Romero_Vila_2008}:

\begin{multline}
    \sigma_{\gamma\gamma}(s) = \frac{3\sigma_T}{16}(1 - \beta_{\rm cm}^2) \\
    \times \left[ (3 - \beta_{\rm cm}^4)\ln\left(\frac{1 + \beta_{\rm cm}}{1 - \beta_{\rm cm}}\right) - 2\beta_{\rm cm}(2 - \beta_{\rm cm}^2) \right]
\end{multline}

This cross-section is non-zero only when the center-of-mass energy squared, $s = \epsilon E_\gamma$, exceeds the threshold for pair production, $(2m_e c^2)^2$. The term $\beta_{\rm cm}$ is defined as

\begin{equation}
    \beta_{\rm cm} = \sqrt{1 - \frac{4 m_e^2 c^4}{s}}
\end{equation}
where $m_e$ is the electron mass, $c$ is the speed of light, $\sigma_T$ is the Thomson cross-section, and $\beta_{\rm cm}$
is the velocity of the created pair in the center-of-mass frame.

In our analysis, we calculate $\tau_{\gamma\gamma}$ by considering the two 
target-photon fields described before: the X-ray coronal photon field and the blackbody radiation field from the 
accretion disk. The X-ray emission was calculated over the corona height $R_{\rm ph}=L$, and the disk blackbody radiation was also calculated considering the inner region contribution (around $R_X$) with temperature $T_d \sim 10^{5}\ \,K$, within $L$. The radiation fields modeled can be seen in the \textit{middle panel} of Figure~\ref{fig:tau_gg}.  

We display the optical depth for each field, and the sum of both (purple line) as a function of the gamma-ray energy in the \textit{top panel} of Figure~\ref{fig:tau_gg}. The luminosity of $\gamma $-rays that escape the emission region will be attenuated by a factor $e^{- \tau_{\gamma\gamma} (E_{\gamma })}$. Hence 
both radiation fields determine the high opacity ($\tau_{\gamma\gamma}$) responsible for suppressing the accompanying TeV $\gamma$-ray emission.

The \textit{bottom panel} of Figure~\ref{fig:tau_gg} depicts the power spectrum of the accelerated  protons.
The proton population is injected with a power-law spectrum of index $\alpha_p= 1.7$ (see Table~\ref{table_parameters} and Section~\ref{subsec:2.4_powerspectrum_partic}). 

\subsection{Particle Acceleration \& Radiative Energy Loss Timescales}
\label{subsec:3.3_pa_radloss_time}

\begin{figure}[htbp]
    \centering
    \includegraphics[width=\linewidth]{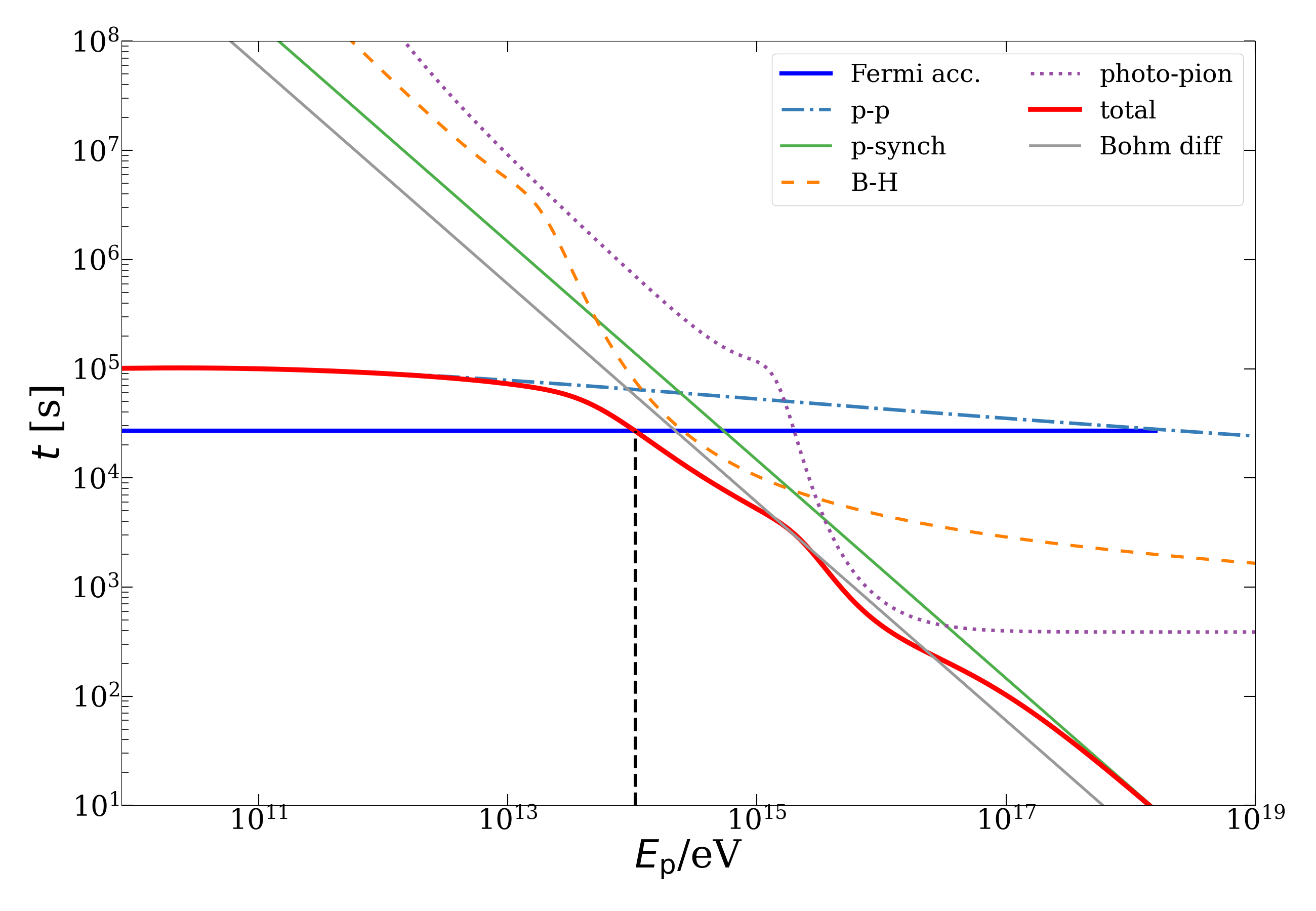}
    \caption{Hadronic Acceleration and Radiative Loss Timescales. The turbulence-driven first-order Fermi acceleration timescale   
    (\textit{Fermi acc.}; solid blue line) is constant and shown up to $E_{\rm p,th}\sim 10^{18}\, \text{eV}$.
    Energy loss timescales include synchrotron radiation (solid green), p-p interactions (blue dot-dashed), photo-meson (\textit{photo-pion}; dotted purple) and \textit{Bethe-Heitler} processes (\textit{B-H}; dashed orange), calculated considering the combined radiation fields from the disk (blackbody) and X-ray corona
    (see Fig.~\ref{fig:tau_gg}). The maximum proton energy, $E_{\rm p,max} \simeq 1.06 \times 10^{14}\, \text{eV}$, is determined by the intersection between the acceleration and total energy-loss timescales, and is indicated by the vertical dashed black line.}
    \label{fig:cool_HAD}
\end{figure}

The energetic balance between particle acceleration and energy loss
is illustrated in~Figure~\ref{fig:cool_HAD} using the parameters derived from observations (Table~\ref{table_observations}) and our model-dependent assumptions (Table~\ref{table_parameters}), as well as the background radiation fields, described in Section~\ref{subsec:3.2_radfield_gammaabsorption}.

The blue solid line in Figure \ref{fig:cool_HAD} represents the timescale for particle acceleration, which is governed by the Fermi process (Eq. \ref{eq:fermi_acc}). The curve terminates
at $E_{\rm p,th} \simeq 7.41 \times 10^{17}$ eV, corresponding to the condition in which the Larmor radius ($r_{L}$) becomes of the order of the thickness of the reconnection sheet ($\Delta R_{X}$ in Eq. \ref{eq:width_rec}), i.e., $r_{L} = 33.36 \text{km} \left( \frac{E}{\text{GeV}} \right) \left( \frac{1}{Z} \right) \left( \frac{\text{G}}{B} \right) = \Delta R_{X}$. Above this energy threshold ($E_{\rm p,th})$, drift acceleration would take place, as described in Section~\ref{subsec:2.2_acc_timescale}.

The primary objective is to determine the maximum achievable proton energy ($E_{p,\max}$) by balancing the acceleration rate with the relevant energy-loss timescales ($t_{\rm loss}$) (Eq.~\ref{loss-rate}). The dominant cooling channels include proton--proton ($t_{pp}$), synchrotron ($t_{\rm syn}$), and photo-hadronic ($t_{p\gamma}$, $t_{\rm BH}$) interactions, presented in the Appendix~\ref{sec:ap_timescales}. For the photo-hadronic processes, the ambient photon field is modeled using the disk blackbody (Optical--UV proxy) and the X-ray component constrained by observations (described in Section~\ref{subsec:3.2_radfield_gammaabsorption}). As shown in Figure \ref{fig:cool_HAD}, the Bethe--Heitler timescale (dashed orange curve) dominates the losses at high energies. The resulting energy balance indicates that the intersection of the acceleration timescale 
with the total loss timescale yields a maximum proton energy of $E_{p,\max} \simeq 1.06 \times 10^{14} \, \text{eV}$.  Above this threshold, protons lose energy faster than they can be further accelerated.

\subsection{The Spectral Energy Distribution of NGC 1068}
\label{subsec:3.4_results}

\begin{figure*}[htbp]
    \centering
    \includegraphics[width=\textwidth]{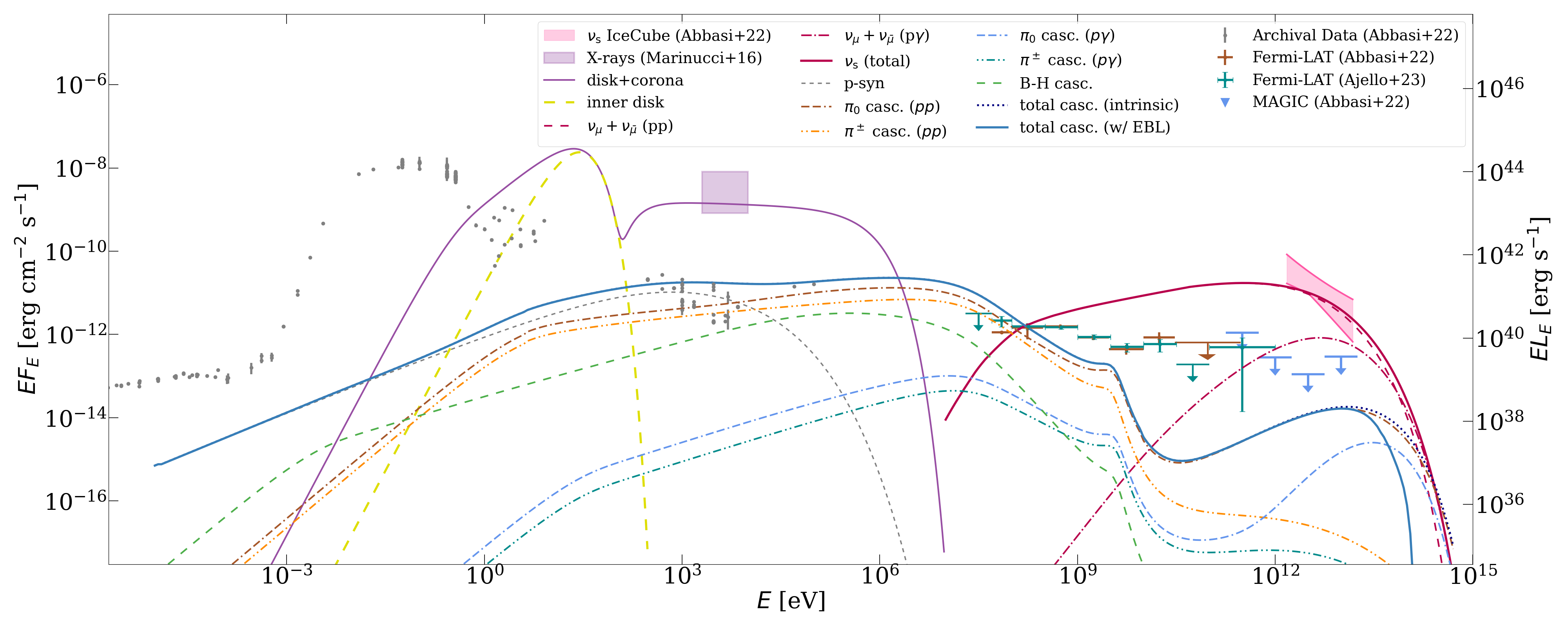}
    \caption{Multi-messenger spectral energy distribution (SED) of the source NGC 1068, derived from the turbulence-driven reconnection acceleration model
    described in this paper. 
    The total neutrino flux (solid magenta, $\nu_\mu$), normalized to represent only the muon neutrino flux, accounts for both $pp$ (dashed magenta) and $p\gamma$ (dot-dashed magenta) interactions. The electromagnetic (EM) output (solid blue) incorporates attenuation by the extragalactic background light (EBL) following \citet{Franceschini2017}; for comparison, the intrinsic EM emission without EBL is shown as the dotted navy curve. These outputs result from self-consistent lepto-hadronic cascades (dotted, dot-dashed, and dashed light blue curves)
    The model highlights heavy $\gamma\gamma$ suppression in the GeV-TeV range due to the coronal X-ray field (solid purple, constrained by \citealt{Marinucci2016_nuStar} shaded purple region) and accretion disk radiation (dashed yellow).
    Observed data were mainly sourced from \citealt{IceCube2022}, which include the IceCube 95\% C.L. neutrino excess (light magenta shaded region),
    archival multi-wavelength data (gray points), and the MAGIC $\gamma$-ray upper limits (blue arrows) \citep{IceCube2022}. Fermi-LAT detections (dark green points; sourced from \citet{Ajello_2023}), are likely produced in extended regions, 
    beyond the compact corona modeled here. 
    }
    \label{fig:SED}
\end{figure*}

We present the modeled multi-messenger Spectral Energy Distribution (SED) of the source NGC 1068 
in Figure~\ref{fig:SED}. This result is obtained by applying the coronal accretion-disk framework described in Section~\ref{sec:2_model} using the parametric space of Tables~\ref{table_observations} and \ref{table_parameters}. In this background, which accounts for the turbulence-driven reconnection acceleration and cooling timescales discussed in Section~\ref{subsec:3.3_pa_radloss_time}, we self-consistently computed the lepto-hadronic losses and subsequent cascades over the emission volume $V$, using the algorithm developed by \citealt{rodriguez_ramirez_etal2020} \citep[see also][]{Rodriguez_Ramirez_2021, dalpino_rodriguez_txs_2025}.

To account for the attenuation of high-energy photons during their propagation through the intergalactic medium, we include
the effect of the Extragalactic Background Light (EBL). For this work, we adopt a redshift of $z = 0.003$ (consistent with Table~\ref{table_observations}) and use the EBL model from \citet{Franceschini2017}. We implemented this by interpolating the tabulated optical depth values $\tau_{\rm EBL}$ onto our electromagnetic cascade grid and applying the absorption multiplicative factor $e^{- \tau_{\rm EBL} (z, E_{\gamma })}$. This correction is particularly significant for the total cascade component at energies above $100$ GeV, ensuring a consistent comparison with the upper limits provided by MAGIC in Fig.~\ref{fig:SED}.



Figure~\ref{fig:SED} summarizes the core result of our model: a turbulence-driven magnetic reconnection region in the inner disk--corona is able to 
reproduce the observed high-energy neutrino flux from NGC~1068, while remaining consistent with the absence of a TeV $\gamma$-ray counterpart. Highlighted by the pink curve, the total muon-neutrino component (solid magenta curve) receives contributions from both $p$--$p$ (dashed magenta) and $p\gamma$ (dot-dashed magenta) interactions. 
This indicates that the neutrino yield is not produced by a single hadronic channel, but rather by the combined effect of dense coronal matter and intense photon backgrounds, with  the $p$--$p$ channel providing the dominant contribution.

In parallel, the electromagnetic cascade emission (blue curves) is strongly suppressed above the GeV range. We note that
the locally produced high-energy photons are efficiently absorbed through $\gamma\gamma$ pair production in the disk and coronal radiation fields.
This is consistent with the large optical depths shown earlier ($\tau_{\gamma\gamma} \gg 1$ in Figure~\ref{fig:tau_gg}) and explains why the source is neutrino-bright but TeV-dim. 
Our findings  also indicate 
that the Fermi-LAT detection points are not likely 
attributed to the  compact corona in the nuclear region alone, supporting the interpretation that the GeV emission arises in more extended regions, such us outflows, circumnuclear starburst, or larger-scale AGN structures \citep[e.g.,][]{Peretti_2023}, which are outside the scope of our compact coronal model.

\subsection{Sweeping the Parametric Space}

\begin{figure*}[htbp]
    \centering 
   \begin{minipage}{0.85\textwidth}
       \centering
       \includegraphics[width=0.85\linewidth]{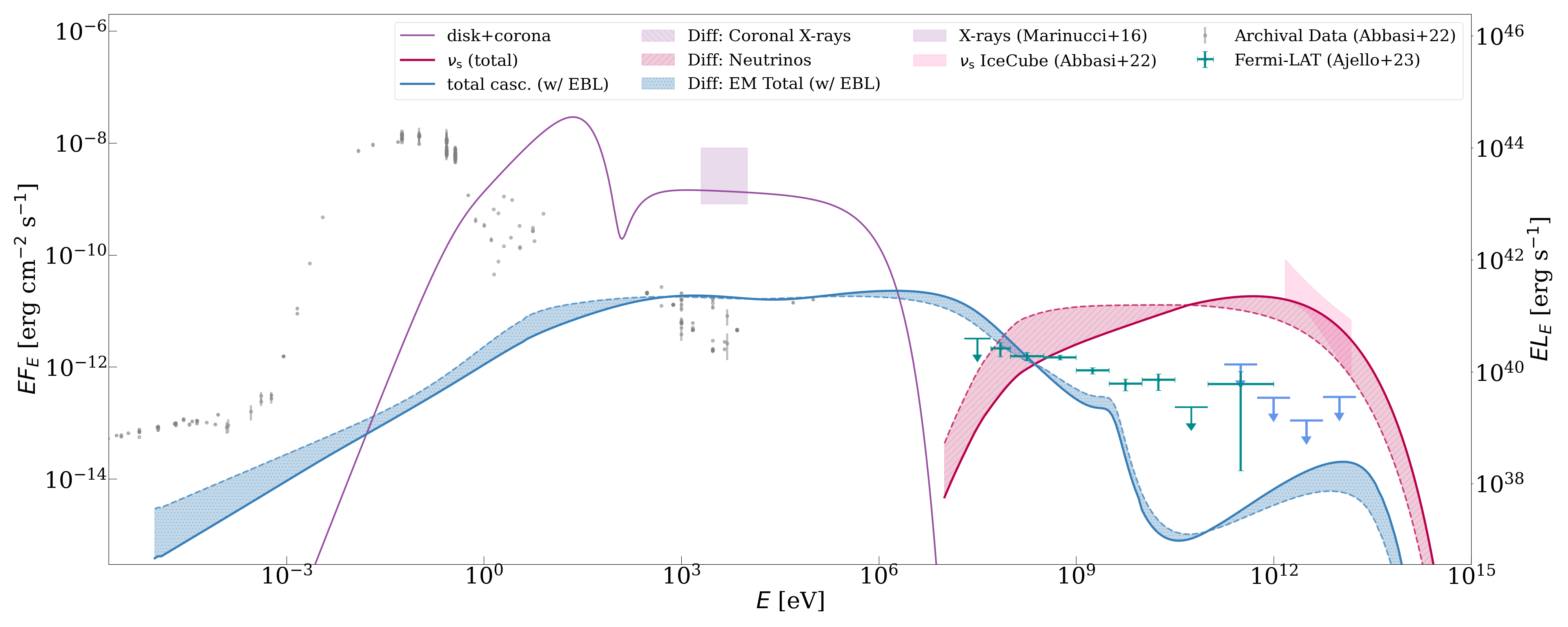}
       \vfill
       \vspace{0.4cm}
   \end{minipage}
    \begin{minipage}{0.85\textwidth}
       \centering
       \includegraphics[width=0.85\linewidth]{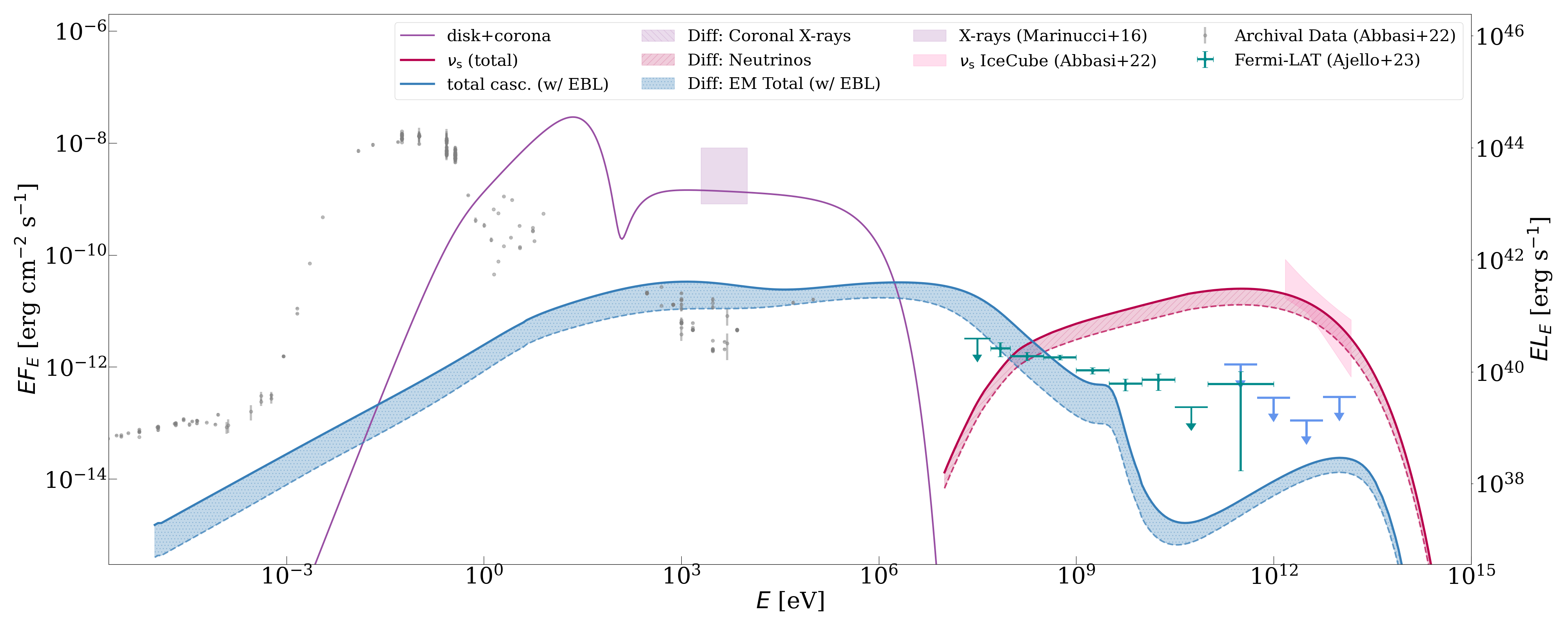}
       \vfill
       \vspace{0.2cm}
    \end{minipage}
    \caption{
    Parameter survey  
    of the NGC 1068 multi-messenger spectral energy distribution (SED). Shaded regions represent the envelope of solutions obtained by varying: the injection proton spectrum slope 
    (upper panel) from $\alpha_{\rm p}=1.6$ (solid curves) to $\alpha_{\rm p}=2.0$ (dashed curves);  and the system geometry ($L_{X}$,$L$) (lower panel), with $L_X = L = 20 \,R_{\rm Sch}$ (dashed curves) to $50 \,R_{\rm Sch}$ (solid curves). 
    The other $free$ parameters are kept the same as those in Table~\ref{table_parameters}.
    All modeled curves maintain the same color-coding as in Figure~\ref{fig:SED}}.
    \label{fig:SED_comparison_alpha_geom}
\end{figure*}

\begin{figure*}[htbp]
    \centering 
   \begin{minipage}{0.85\textwidth}
       \centering
       \includegraphics[width=0.85\linewidth]{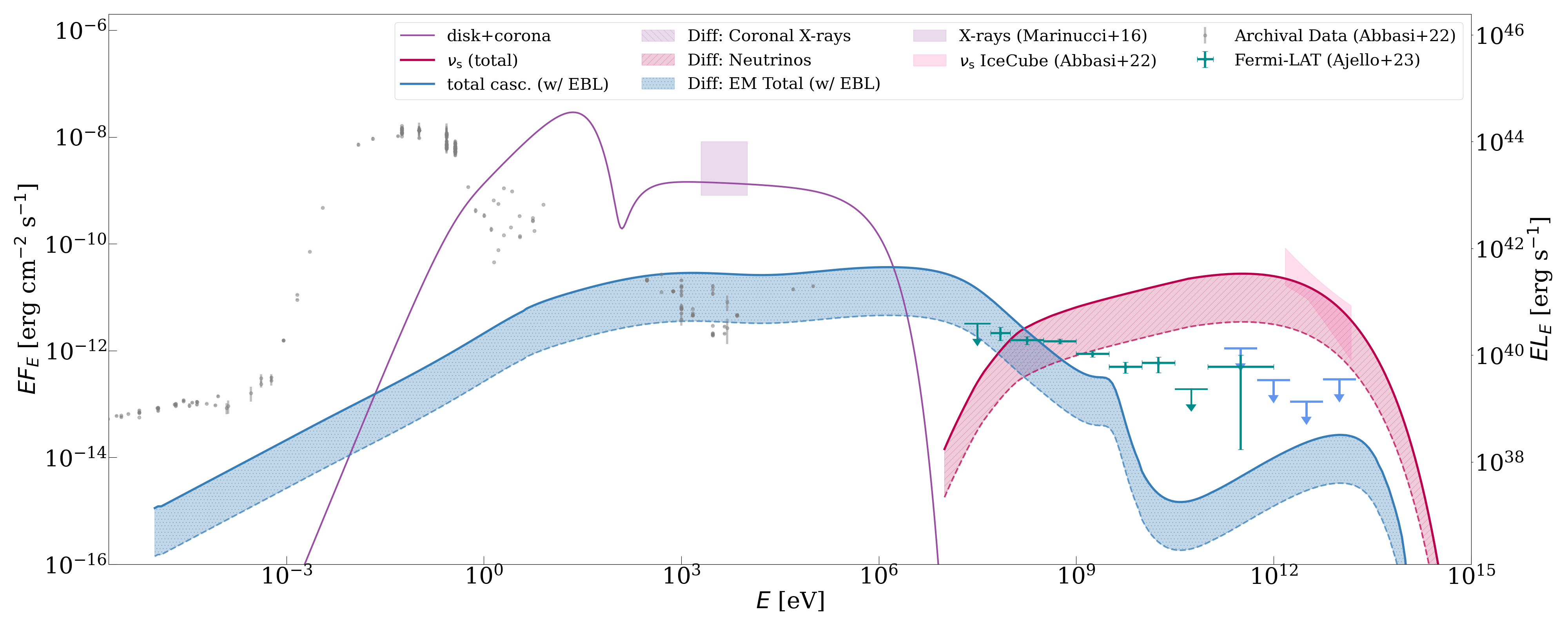}
       \vfill
       \vspace{0.2cm}
   \end{minipage}
    \begin{minipage}{0.85\textwidth}
       \centering
       \includegraphics[width=0.85\linewidth]{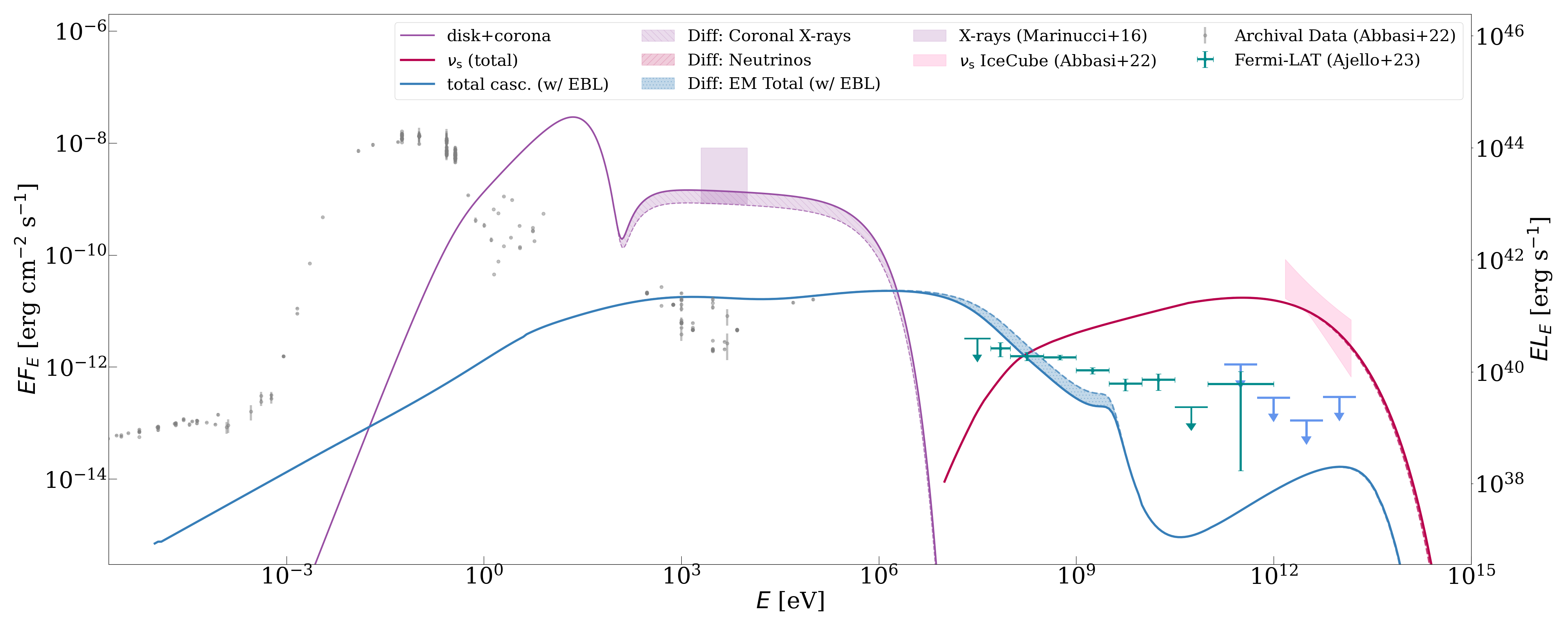}
       \vfill
       \vspace{0.2cm}
    \end{minipage}
    \caption{Parameter survey of the NGC 1068 multi-messenger spectral energy distribution (SED). Shaded regions represent the envelope of solutions obtained by varying:  
    the fraction  $\eta_{\rm p}$ of magnetic reconnection power injected into protons, ranging from $\eta_{\rm p}=$ 0.1 to 0.8 (upper panel), and the efficiency of accretion power $\eta_{\rm cx}$ converted into 
    X-ray luminosity, ranging from $\eta_{\rm cx} = 0.005-0.0085$
    (lower panel). The other $free$ parameters are  the same as those in Table~\ref{table_parameters}. All modeled curves maintain the same color-coding as in Figure~\ref{fig:SED}.}
    \label{fig:SED_comparison_etas}
\end{figure*}

Figure~\ref{fig:SED} shows our best-fit model for the fiducial set of free parameters listed in Table~\ref{table_parameters}. Although this fiducial configuration provides the reference solution, no spectral fitting procedure was performed, and the parameters are not uniquely fixed, allowing a certain 
variation within physically reliable ranges for this source. To illustrate this, the SEDs shown in Figures~\ref{fig:SED_comparison_alpha_geom} and \ref{fig:SED_comparison_etas} explore the impact of varying selected $free$ parameters around their fiducial values (in Table~\ref{table_parameters}).

Figures~\ref{fig:SED_comparison_alpha_geom} and \ref{fig:SED_comparison_etas} reveal that the model admits a family of viable solutions with controlled parameter degeneracies, not restricted to a single fine-tuned configuration. 

The upper panel of Figure~\ref{fig:SED_comparison_alpha_geom}, for instance,  shows that the neutrino fit is relatively insensitive to the proton injection slope for $\alpha_p \simeq 1.6$--$1.7$, while spectra approaching $\alpha_p \sim 2$ become too soft and underproduce the IceCube excess at the highest energies. Not shown in the figure, we find that values $\alpha_p \simeq 1.8$--$1.9$ remain consistent with IceCube data, if we consider higher efficiencies such as $\eta_p\sim 0.6-0.7$, or larger geometries.

The lower panel of Figure~\ref{fig:SED_comparison_alpha_geom} highlights an important geometry--efficiency degeneracy: smaller coronal and reconnection scales require larger $\eta_p$ to reach the same neutrino luminosity, whereas larger coronal structures can match the data with more moderate efficiencies. This is physically reasonable, since increasing the size $(l_X,l)$ increases the magnetic energy available for reconnection and therefore the total hadronic power. The values of $l_X = l$ in the figure range from $20 \,R_{\rm Sch}$ to $50 \,R_{\rm Sch}$. Table \ref{table_appendix} in Appendix \ref{sec:ap_param_space_table} shows the set of assumed and model-derived  parameters for this family. We also tested configurations with $l_X < l$ over the same interval and found qualitatively similar results.

The top panel of Figure~\ref{fig:SED_comparison_etas} shows that the model is only mildly sensitive to variations in the fraction of turbulence-driven magnetic reconnection power channeled into proton acceleration over the range $\eta_{\rm p}=0.1$--$0.8$. 
Although the neutrino flux remains consistent with the data for the larger value $\eta_{\rm p}=0.8$, the corresponding increase in the predicted \textit{Fermi}-band $\gamma$-ray emission leads to greater tension with the observed flux, suggesting that, for higher reconnection power, the coronal region could also account for part of the emission in this energy range.

Finally, the lower panel of Figure~\ref{fig:SED_comparison_etas} illustrates the impact of varying the X-ray luminosity by changing the fraction $\eta_{\rm cx}$ of the accretion power converted into X-ray emission, $L_{\rm X}\simeq \eta_{\rm cx}\dot{M}c^{2}$. For $\eta_{\rm cx}=0.005$--$0.0085$, we obtain $L_{\mathrm{X\text{-}ray}}=(6.9\times10^{43}$--$1.2\times10^{44})\,\mathrm{erg\,s^{-1}}$, in agreement with the observed range \citep{Marinucci2016_nuStar, Bauer_2015_nuStar}.
In this sense, Figures~\ref{fig:SED_comparison_alpha_geom} and \ref{fig:SED_comparison_etas} strengthen the robustness of the scenario while also making explicit which parameter combinations are observationally degenerate.

\section{Summary and Discussion}
\label{sec:4_discussion}

In this work 
we have modeled the hadronic components
of the spectral energy distribution of the active galaxy NGC1068, by using a fraction of the magnetic reconnection power, potentially released within the nuclear region, to accelerate protons near the BH horizon. Based upon earlier work  \citep{dalpino_lazarian_2005, dalpino_etal_2010_GPK10,kadowaki_etal_15}, we revisited their framework in order to use turbulence-driven fast magnetic reconnection  as the main mechanism operating in the inner region of the Seyfert Type II galaxy NGC1068. 


We find that protons accelerated via  first-order Fermi process within the turbulent reconnection layer, formed by the interaction between magnetic field lines rising from the inner accretion disk and poloidal field lines anchored to the central black hole horizon (see Figure~\ref{fig:coronal_model}), experience a local magnetic field strength of $B_c \simeq B \simeq B_d \sim 1.8 \times 10^{4}$ G, for the adopted parameters $r_X=6$, $l_X=30$, $l=30$, $\alpha=0.1$.
The total magnetic power released by this process is estimated to be of order $\sim 10^{43}$ erg s$^{-1}$. With roughly $50\%$ of this power injected into protons, it is sufficient to reproduce the neutrino flux observed by IceCube \citep{IceCube2022, IceCube2025}. The resulting  neutrino emission arises from both $p$–$p$ and $p\gamma$ interactions, with the $p$–$p$ channel dominating.
Remarkably, 
the reconnection velocity predicted by turbulence-driven reconnection, for the parameters of NGC 1068, is only a tiny fraction of the local Alfvén speed, implying that the observed neutrino flux does not require extreme flaring episodes. Instead, it can be sustained by near steady-state reconnection, given the balance between the particle acceleration time (of the order of the particle escape time) and
the timescale for particle losses over the characteristic dimensions of the reconnection region ($\Delta R_X$, $L_X$).

The model naturally explains the lack of a $\gamma$-ray counterpart by showing strong absorption in the TeV-PeV range. While high-energy $\gamma$-rays are produced alongside neutrinos, via neutral pion decay from proton-proton ($pp \rightarrow \pi^0$) and photo-meson ($p\gamma$ $\rightarrow \pi^0$) interactions, they  anihilate with low-energy background photons via pair production ($\gamma\gamma \rightarrow e^\pm$), effectively explaining the "hidden" nature of the neutrino source \citep{Murase2022}. The dense radiation environment presents a high optical depth ($\tau_{\gamma\gamma} \gg 1$), as illustrated in the top panel of Figure~\ref{fig:tau_gg}, dominated by the OUV blackbody disk radiation and X-ray coronal field, consistent with Chandra and NuSTAR data \citep{Bauer_2015_nuStar, Marinucci2016_nuStar}. We note that the X-ray emission corresponds to a phase of reduced obscuration and displays the variability typical of this source \citep{Bauer_2015_nuStar,IceCube2025}. 
We refer to \citet{PassosReis_NGC_ICRC2025}  for a glimpse of how this variability may impact the  multi-messenger SED  of NGC 1068.

Our exploration of the parametric space demonstrates that the proposed scenario remains viable across changes of the proton injection spectral index, the fraction of reconnection power channelled into protons ($\eta_p$), the coronal geometry, and the efficiency with which accretion power emerges as X-rays. In particular, the model exhibits a clear degeneracy between source size and reconnection power efficiency: more compact configurations can still match the IceCube data, but only if a larger fraction of the reconnection power is transferred to relativistic protons, whereas more extended coronal geometries require more moderate efficiencies. This behavior is physically consistent with the scaling of reconnection power with the size of the emitting region, and it suggests that future constraints on the coronal geometry (e.g., from X-ray reverberation or variability studies) could directly narrow the allowed range of $\eta_p$.

Overall, our results show that the neutrino flux is primarily regulated by the total magnetic reconnection power available in the inner corona, while the spectral shape is largely determined by the acceleration mechanism, the spectral index of particle injection $\alpha_\mathrm{p}$, and radiative losses. Variations in geometry mainly rescale the energetic normalization of the emission, whereas variations in $\alpha_p$ modify the spectral slope at the highest energies. The efficiency parameter $\eta_p$ then compensates these effects by adjusting the fraction of magnetic power transferred to nonthermal protons, here assumed to be around $50\%$. The variation in $\eta_{\rm cx}$ mimics the variability observed in the X-ray band. This separation of roles 
clarifies which parameters control normalization and which control spectral shape. Hence, the existence of multiple viable parameter combinations indicates that the neutrino production is a robust outcome of the reconnection scenario rather than the result of parameter fine tuning.


Even our best-fitting model hints at a possible “small contamination” of $\gamma$-rays at the lowest Fermi-LAT energies (Figure~\ref{fig:SED}). This mild tension persists throughout the full parameter scan. In this work we adopted solutions with $r_X \simeq 6$ to avoid complications from strong general-relativistic effects near the BH horizon; however, smaller radii would lead to nearly complete absorption in the Fermi-LAT band, further obscuring the core, as suggested by preliminary results \citep[see][]{PassosReis_NGC_ICRC2025}.

The results of the present work confirm that the neutrino excess observed by IceCube  \citep{IceCube2022, IceCube2025}, can originate in the  vicinity of the BH horizon, without violating the upper limits set by MAGIC \citep{Acciari2019_MAGIC} and Fermi-LAT \citep{Ajello_2023}.

The residual $\gamma$-ray emission observed from NGC 1068 is likely not  emission from the core, but rather produced in the extended circumnuclear environment in AGN outflows such as jets, starburst activity, or ultra-fast outflows (UFOs) \citep{Cecil1990_outflows, Romeo2016_starburst, Eichmann2016_starburst, Salvatore2024_jet, Peretti_2023}. 



Unlike previous models that rely on plasmoid-driven reconnection, external shock acceleration and/or external pre-acceleration zones (Section \ref{sec:1_intro}), our approach considers turbulence-driven reconnection located in a single zone in the inner region of the corona to accelerate particles in a Fermi process. 
It effectively reproduces the observed neutrino excess 
with negligible contribution to the observed gamma-ray emission of the source. 

In summary, our  results support turbulence-driven magnetic reconnection as a viable and efficient mechanism for hadronic acceleration and neutrino production in the inner corona of NGC~1068. The model simultaneously accounts for
three observational facts:
(i) the IceCube neutrino excess, (ii) the absence 
of a TeV electromagnetic counterpart, and (iii) the likely non-coronal origin of the observed GeV gamma-ray emission. This makes NGC~1068 a compelling prototype for a broader class of obscured 
AGN, in which  compact magnetized coronae may act as hidden neutrino factories, with the electromagnetic signature largely suppressed by internal photon fields.

Although the present framework is successful, several extensions are important. First, the model is steady-state, while NGC~1068 shows X-ray obscuration variability; time-dependent modeling could test how changes in the coronal photon density affect both the neutrino output and the $\gamma\gamma$ opacity. Second, the compact-corona treatment is intentionally one-zone; a multi-zone implementation could better connect the coronal neutrino source to the larger-scale GeV-emitting environment. Third, solutions at smaller radii ($r_X <6$) likely remain promising for NGC 1068 and other Seyfert galaxies, though they require relativistic corrections to the central potential approximation, a refinement being detailed in forthcoming work, which improves upon the original \citealt{dalpino_lazarian_2005} model.
These improvements would strengthen the predictive power of the model and help assess whether the same mechanism can explain neutrino emission in other Seyfert galaxies.

\begin{acknowledgments}
LPR and EMdGDP acknowledge the support from the Brazilian Funding Agency FAPESP (grants n. 2021/02120-0; 2024/05459-6; 2020/11891-7; 2023/10590-1). EMdGDP also acknowledges support from CNPq (grant 308643/2017-8).

\end{acknowledgments}

\section*{Data Availability}
The data of this article will be available upon request to the authors.

%



\software{ 
          Python3 \citep{python3}
          }

\newpage


\appendix

\section{Characteristic Timescales for Particle Interactions}
\label{sec:ap_timescales}

The maximum energy of accelerated protons ($E_{\rm p,max}$) is determined by the competition between the first-order Fermi acceleration timescale ($t_{\rm acc, fermi}$, Eq.~\ref{eq:fermi_acc}) and the various energy loss timescales ($t_{\rm loss}$) in the corona, as illustrated in Figure \ref{fig:cool_HAD}. The total loss rate is given by the sum of the inverse timescales for synchrotron, proton-proton, photo-hadronic (photopion and Bethe-Heitler) processes:

\begin{equation}
    \frac{1}{t_{\rm loss}(E_p)} = \frac{1}{t_{\rm sync}(E_p)} + \frac{1}{t_{\rm pp}(E_p)} + \frac{1}{t_{\rm p\gamma }(E_p)} + \frac{1}{t_{\rm BH}(E_p)}
\end{equation}

The specific formulations for the key cooling timescales used in the model are presented below.

\subsection{Synchrotron Cooling ($t_{\rm sync}$)}

The timescale for a charged particle $i$ (proton or electron/positron) to lose energy $E_i$ via synchrotron radiation in a magnetic field with energy density $U_B = B^2 / (8\pi)$ is given by:

$$t_{\rm sync}^{-1} (E_{i}) = \frac{P_{\rm syn, tot}}{E_{i}} = \frac{4}{3} \left( \frac{m_{e}}{m_{i}} \right)^{2} \sigma_{T} c\ U_{B} \frac{E_{i}}{(m_{i} c^{2})^{2}}$$

where $\sigma_{T}$ is the Thomson cross-section and $m_e$ and $m_i$ are the rest masses of the electron and the radiating particle, respectively.

\subsection{Proton-Proton Collisions ($t_{\rm pp}$)}

Proton-proton (p-p) interactions refer to collisions between high-energy protons and the ambient matter. These interactions produce secondary particles, including neutrinos, gamma-rays, and pions, through inelastic collisions.

The cooling rate for p-p interactions can expressed as \cite{khiali_etal_15}:

\begin{equation}
    t_{\rm pp}^{-1} = n_{i} c \sigma_{\rm pp} k_{\rm pp},
\end{equation}

where $k_{\rm pp} \approx 0.5$ is the inelasticity of the process, and the corresponding cross-section for inelastic p-p interactions, $\sigma_{\rm pp}(E_p)$, is approximated by \citet{Kelner_Aharonian_Bugayov_2006, Kelner_Aharonian_Bugayov_2009}:

\begin{equation}
    \sigma_{\rm pp} (E_{p}) = \left ( 34.3 + 1.88 \mathcal{L} + 0.25 \mathcal{L}^{2} \right ) \left [ 1 - \left ( \frac{E_{\rm th}}{E_{p}} \right )^{4} \right ]^{2} \text{ mb}
\end{equation}

with $ 1 \text{ mb} = 1 \times 10^{-27} \text{ cm}^{2}$ and

\begin{equation}
    \mathcal{L} = \ln{\left ( \frac{E_{p}}{1 \text{ TeV}} \right )}.
\end{equation}

The proton threshold kinetic energy or neutral pion ($\pi^{0}$) production is

\begin{equation}
    E_{\rm th} = 2 m_{\pi } c^{2} \left ( 1 + \frac{m_{\pi }}{4 m_{p}} \right ) \approx 280 \text{ MeV},
\end{equation}

where $m_{\pi }c^{2} = 134.97$ MeV is the rest energy of $\pi^{0}$ \citep{VilaAharonian2009}.

\subsection{Photopion Production ($t_{\rm p\gamma }$)}

The photomeson production rate ($p + \gamma \rightarrow p + \pi^0/\pi^{\pm} + \dots$) is a critical loss mechanism for protons interacting with the ambient photon field $n_{\rm ph} (E_{\rm ph})$.

\citet{Atoyan2003} proposed a simplified approach to calculate the cross-section and the inelasticity \citep[see also][]{Dermer_Atoyan_2003} which are given by

\begin{equation}
    \sigma_{p \gamma } \approx \left\{ \begin{matrix}
        340 \ \mu barn & \quad 300 MeV \leq \epsilon_{r} \leq 500 MeV \\
        120 \ \mu barn & \quad \epsilon_{r} > 500 MeV,
    \end{matrix} \right.
\end{equation}

and

\begin{equation}
    K_{p\gamma } \approx \left\{ \begin{matrix}
        0.2 & \quad 300 MeV \leq \epsilon_{r} \leq 500 MeV \\
        0.6 & \quad \epsilon_{r} > 500 MeV.
    \end{matrix} \right.
\end{equation}

The photomeson production takes place for photon energies greater than $E_{\rm th} \approx 145$ MeV. A single pion can be produced in an interaction near the threshold and then decay giving rise to $\gamma $-rays.

When high-energy protons interact with low-energy photons in radiation fields, photomeson production occurs, leading to the generation of neutrinos and gamma-rays. This process is particularly important in environments rich in ultraviolet (UV) and X-ray photons. The energy of protons in the photon field is:

\begin{equation}
    \gamma_{p} = \frac{E_{p}}{m_{e} c^{2}},
\end{equation}

$\epsilon_{r}$ is the rest frame of the proton and $K_{p\gamma }^{(\pi )}$ is the inelasticity of the interaction, where $n_{ph} (E_{ph})$ is the photon distribution, and $\sigma_{\rm p\gamma }^{(\pi )} $ is the cross-section for the interaction.

Then, the inverse timescale for photomeson production is calculated by integrating over the photon energy $E_{\rm ph}$ and the proton rest-frame energy $\epsilon_r$:


\begin{equation}
    t_{\rm p\gamma }^{-1} (E_{p}) = \frac{c}{2\gamma_{p}^{2}} \int_{\frac{E_{\rm th}^{(\pi )}}{2 \gamma_{p}}}^{\infty } \mathrm{d} E_{\rm ph} \frac{n_{\rm ph} (E_{\rm ph})}{E_{\rm ph}^{2}} \\ \int_{E_{\rm ph}^{(\pi )}}^{2 E_{\rm ph} \gamma_{p}} \mathrm{d} \epsilon_{r}\ \sigma_{\rm p\gamma }^{(\pi )} (\epsilon_{r}) K_{\rm p\gamma }^{(\pi )} (\epsilon_{r})\ \epsilon_{r}.
\end{equation}

where $E_{\rm th}^{(\pi )}$ is the threshold energy for photopion production.

\subsection{Bethe-Heitler Pair Production ($t_{\rm BH}$)}

The Bethe-Heitler process involves the production of electron-positron pairs through the interaction of high-energy protons with radiation fields ($p + \gamma \rightarrow p + e^{+} + e^{-}$). This cooling mechanism is a lower-threshold photo-hadronic interaction, becoming important when protons interact with photons, such as those in the UV or X-ray band. The inverse timescale is approximated as:


\begin{equation}
    t_{\rm BH}^{-1} (\gamma_p) \approx \frac{c}{\gamma_{p}^{2}} \int_{\bar{\epsilon}_{\rm BH} / (2 \gamma_{p} )}^{\infty} d\epsilon \frac{dn_{x}}{d\epsilon } \epsilon^{-2} \int_{\bar{\epsilon}_{\rm BH}}^{2\gamma_{p} \epsilon} \epsilon^{'} \sigma_{\rm p\gamma }^{(\pi )}(\epsilon^{'}) d\epsilon^{'};
\end{equation}

with $\bar{\epsilon}_{\rm BH} = 2 MeV $, as the threshold energy for pair production in the proton rest frame.

\section{Geometrical Parameters Survey}
\label{sec:ap_param_space_table}

In this appendix we present, as an example, the extended exploration of the geometrical parameters adopted in our magnetic reconnection model.
Table~\ref{table_appendix} reports the representative lower and upper bound solutions  used in the lower panel of Figure \ref{fig:SED_comparison_alpha_geom}, where we changed the geometrical parameters ($l, l_X$).  The model admits a family of acceptable solutions. 
The  derived quantities in the table follow directly from these geometric scalings. In particular, the width of the current sheet and the total reconnection power vary proportionally to the size of the acceleration region, while local plasma conditions (temperature and density) remain nearly unchanged. This demonstrates that the geometry mainly controls the energetic normalization of the emission rather than its spectral properties (see more details in Section \ref{sec:3_results}).

\begin{table*}[htbp]
    \centering
    \caption{Parameter bounds for the hatched regions in the spectral energy distribution (SED). The columns show the lower and upper limits adopted for the modeled curves of the bottom panel of Fig.~\ref{fig:SED_comparison_alpha_geom}. Note that  all  free parameters remained unchanged from Table~\ref{table_parameters}, except the pair $(l,l_X)$.
    }
    \begin{tabular}{lcccc}
    \hline
    \hline
    This Work & Parameter & Lower Fit & Upper Fit & Unit \\
    \hline
    Coronal magnetic flux tube height & $l$ & 20 & 50 & [$L/R_{\rm Sch}$] \\
    Height of reconnection region & $l_X$ & 20 & 50 & [$L_X/R_{\rm Sch}$] \\
    Inner radius of disk & $r_X$ & 6 & 6 & [$R_X/R_{\rm Sch}$] \\
    Spectral index of the injected proton distribution & $\alpha_{\rm p}$ & 1.7 & 1.7 \\
    Accretion disk viscosity  & $\alpha $ & 0.1 & 0.1 & \\
    Accretion power conversion efficiency to coronal X-ray luminosity  & $\eta_{\rm cx} $ & $0.0085$ & $0.0085$ & \\
    Fraction of reconnection power transferred to protons & $\eta_{\rm p} $ & $0.5$ & $0.5$ & \\
    \\
    \hline
    \hline
    Model-derived Parameters & & & & \\
    \hline
    Coronal magnetic field & $B_c$ & $1.8 \times 10^{4}$ & $1.8 \times 10^{4}$ & [G] \\
    Coronal particle density & $n_c$ & $2.9 \times 10^{10}$ & $1.4 \times 10^{10}$ & [cm$^{-3}$] \\
    Coronal temperature & $T_c$ & $3.3 \times 10^{9}$ & $3.7 \times 10^{9}$ & [K] \\
    Surface Disk temperature & $T_d$ & $1.5 \times 10^{5}$ & $1.5 \times 10^{5}$ & [K] \\
    Width of current sheet & $\Delta R_{X}$ & $1.3 \times 10^{11}$ & $1.8 \times 10^{11}$ & [cm] \\
    \quad \quad in Schwarzschild radii & & $=0.022 \,R_{\rm Sch}$ & $= 0.031 \,R_{\rm Sch}$ & [$R_{\rm Sch}$] \\
    Reconnection Power released & $\dot{W}_{B}$ & $1.4 \times 10^{43}$ & $2.8 \times 10^{43}$ & [erg\,s$^{-1}$] \\
    Reconnection velocity & $v_{\rm rec}$ & 0.001 & 0.0006 & [$v_{\rm A}$] \\
    Proton power ($\eta_{\rm p} \dot{W}_{B}$) & $L_{p}$ & $7.2 \times 10^{42}$ & $1.4 \times 10^{43}$ & [erg s$^{-1}$] \\
    \hline
    \hline
    \end{tabular}
    \label{table_appendix}
\end{table*}







\bibliography{sample631}{}
\bibliographystyle{aasjournal}




\end{document}